\documentclass[%
 reprint,
nofootinbib,
 amsmath,amssymb,
 aps,
]{revtex4-1}

\usepackage{graphicx}
\usepackage{dcolumn}
\usepackage{bm}

\begin{document}


\title{Ensemble fluctuations of the cosmic ray energy spectrum and the 
intergalactic magnetic field}

\author{A.~D.~Supanitsky}
\affiliation{Instituto de Astronom\'ia y F\'isica del Espacio (IAFE, CONICET-UBA), CC 67, Suc.~28 (C1428ZAA) 
Ciudad Aut\'onoma de Buenos Aires, Argentina}
\email{supanitsky@iafe.uba.ar}

\author{G. Medina-Tanco}
\affiliation{Instituto de Ciencias Nucleares, UNAM, \\ Circuito Exteriror S/N, Ciudad Universitaria,
M\'exico D. F. 04510, M\'exico}

\date{\today}

\begin{abstract}

The origin of the most energetic cosmic ray particles is one of the most important open problems in 
astrophysics. Despite a big experimental effort done in the past years, the sources of these very energetic
particles remain unidentified. Therefore, their distribution on the Universe and even their space density are 
still unknown. It has been shown that different spatial configurations of the sources lead to different energy 
spectra and composition profiles (in the case of sources injecting heavy nuclei) at Earth. These ensemble 
fluctuations are more important at the highest energies because only nearby sources, which are necessarily few, 
can contribute to the flux observed at Earth. This is due to the interaction of the cosmic rays with the low 
energy photons of the radiation field, present in the intergalactic medium, during propagation. It is believed 
that the intergalactic medium is permeated by a turbulent magnetic field. Although at present it is still unknown, 
there are several constraints for its intensity and coherence length obtained from different observational techniques. 
Charged cosmic rays are affected by the intergalactic magnetic field because of the bending of their trajectories 
during propagation through the intergalactic medium. In this work, the influence of the intergalactic magnetic field 
on the ensemble fluctuations is studied. Sources injecting only protons and only iron nuclei are considered. The 
ensemble fluctuations are studied for different values of the density of sources compatible with the constraints 
recently obtained from cosmic ray data. Also, the possible detection of the ensemble fluctuations in the context 
of the future JEM-EUSO mission is discussed.
 
\end{abstract}

\maketitle


\section{Introduction}
\label{Intro}

The origin of the ultrahigh energy cosmic rays (UHECRs) is still unknown. However, big progress on the 
understanding of this phenomenon has been achieved in past years due to the good quality data taken by current 
observatories. It is believed that UHECRs are accelerated in extragalactic objects and propagate through 
the Universe to reach Earth. The main source candidates are active galactic nuclei, radio galaxies, and 
gamma-ray bursts (see, for instance, Ref.~\cite{Kotera:11}). It is also believed that at low energies the cosmic 
rays originate in galactic astrophysical objects like supernova remnants \cite{Hillas:06}. Therefore, the 
galactic-extragalactic transition of the cosmic ray flux should be located in an intermediate energy region.  

The cosmic ray energy spectrum observed at Earth is a very rich source of astrophysical information. It can be 
roughly approximated by a broken power law with four spectral features: the knee at a few $10^{15}$ eV, the ankle 
at $\sim 7\times10^{18}$ eV, the cutoff or suppression at $\sim 4\times10^{19}$ eV, and a second knee at 
$\sim 10^{17}$ eV, recently reported by the KASCADE-Grande Collaboration \cite{KG:11}. The origin of the suppression 
is not clear yet. It can be due to the interaction of the cosmic rays with the radiation fields in the intergalactic 
medium, on the route to Earth, the inefficiency of the sources to accelerate particles at higher energies, or a 
combination of both effects.    

The composition profile as a function of energy is also a valuable piece of information for the understanding of 
the origin of the cosmic rays. In this regard, the data collected by The Pierre Auger Observatory, located in the 
Southern Hemisphere, shows a change that marks the beginning of a transition from light to heavy primaries 
\cite{Auger:10a,Auger:13a}. Such a transition is not confirmed by the Telescope Array\footnote{Note that the Telescope 
Array is located in the Northern Hemisphere.} data, which is more compatible with a proton-dominated composition 
\cite{TA:14a}. However, the present statistics of the Telescope Array data is not enough to distinguish between 
the composition profile seen by Auger and a proton-dominated case \cite{Hanlon:13}. Note, however, that regarding 
composition the interpretation of the experimental data is based on shower simulations for which high energy 
hadronic interaction models, that extrapolate low energy data from accelerators up to the energy of the cosmic rays, 
are used. Therefore, the composition determination is subject to systematic uncertainties coming from such extrapolation 
of the high energy hadronic interactions. In any case, there is a real possibility that nuclei heavier than protons 
dominate the energy spectrum at the highest energies.

At present, there is no identification of any cosmic ray extragalactic source. The data from The Pierre Auger 
Observatory show an excess in the region of Centaurus A \cite{Auger:10}. Also, the data from the Telescope 
Array observatory show a hot spot located close to the supergalactic plane \cite{TA:14}. However, the statistical
significance of both excesses is still quite low, and more data are required to draw any solid conclusion.

The trajectories of the charged cosmic rays are bent by their interaction with the galactic and extragalactic magnetic 
fields. Heavier nuclei are more affected by these magnetic fields. Therefore, the presence of heavy nuclei at the highest 
energies would make much more difficult the identification of the sources. The intergalactic magnetic field (IGMF) is 
poorly known (see Ref.~\cite{Beck:11} for a review). Faraday rotation measurements show that in the core of galaxy 
clusters the magnetic field can take values between 1 and $40\ \mu$G. Outside clusters, upper limits to the magnetic field 
intensity of the parallel component to the line of sight have been obtained by using rotation measurements 
\cite{Blasi:99,Ryu:98}: $\langle B_{||}^2\ L_c \rangle^{1/2} \lesssim 10$ nG Mpc$^{1/2}$, where $L_c$ is the coherence 
length. Recently, a new and very promising technique to constrain the IGMF has been developed. It consists in the study 
of the effects of a non-null IGMF on the propagation of gamma rays originated in blazars. For instance, in Ref.~\cite{HESS:14}, 
it is shown that the application of this new method to the gamma-ray emission coming from the blazar PKS 2155-304, leads to
the exclusion of strengths of the IGMF in the range $(0.3-3)\times 10^{-15}$ G at the 99 \% confidence level, under the assumption 
of a 1 Mpc coherence length. Note that the two methods mentioned above which are used to investigate the IGMF are based on 
physical principles quite different. While the methods based on gamma-ray observations of blazars make use of the broadening 
of the angular distribution of photons and the modifications of the energy spectrum at low energies in the case of blazars 
with hard spectra, the technique based on the observations of Faraday rotation makes use of the rotation of the polarization 
plane of the polarized light emitted by extragalactic objects.   
 
As mentioned before, the sources of UHECR are not identified yet. Therefore, different spatial configurations of the 
sources give different energy spectra and, in the case of heavy nuclei at the sources, different composition profiles
at Earth. These ensemble fluctuations have been studied in detail in Ref.~\cite{Ahlers:13} for the case in which the 
sources inject either only protons or iron nuclei. Moreover, in those studies the IGMF was neglected. In this work, 
we study the influence of a non-null IGMF on the ensemble fluctuations. We also consider pure proton and iron for the 
injected composition, which correspond to the two limiting cases. 

The paper is organized as follow. In Sec.~\ref{Sims} the simulations of the IGMF and the cosmic ray propagation
in the intergalactic medium are described. In Sec.~\ref{EF} the calculations of the ensemble fluctuations for a non-null 
IGMF are presented and discussed. The possibility of its observation with future cosmic ray observatories is discussed in 
Sec.~\ref{POJE} and we conclude in Sec.~\ref{conc}.

\section{Simulations}
\label{Sims}

As mentioned before the intergalactic magnetic field is poorly known. Therefore,
a common approach is to assume that the Universe is filled with a turbulent and
homogeneous magnetic field. In this work, the simulation of the field is done 
following Ref.~\cite{Tautz:13}, where an improved version of the method introduced 
in Ref.~\cite{Giacalone:94} is developed. The magnetic field in every point of the
space $\mathbf{x}$ is given by
\begin{equation}
\label{Bsim}
\mathbf{B}(\mathbf{x}) = \epsilon \sum_{i=1}^{N} {\bm \xi}_i\; A(k_i)%
\cos(\mathbf{k}_i \cdot \mathbf{x}+\zeta_i),
\end{equation}
where $\epsilon=1/\sqrt{2}$, $\zeta_i$ is a random phase angle uniformly distributed in
$[0,2\pi]$, and 
\begin{equation}
\label{k}
\mathbf{k}_i =k_i \left(
\begin{array}{c}
 \sin\theta_i \cos\phi_i    \\
 \sin\theta_i \sin\phi_i    \\
 \cos\theta_i
\end{array}
\right)
\end{equation}
with $\theta_i$ uniformly distributed in $[0,\pi]$ and $\phi_i$ uniformly 
distributed in $[0,2\pi]$. The vector ${\bm \xi}_i$ is chosen orthogonal to 
$\mathbf{k}_i$. It can be written as 
\begin{equation}
\label{xi}
{\bm \xi}_i = \left(
\begin{array}{c}
 -\sin\phi_i \cos\alpha_i + \cos\theta_i \cos\phi_i \sin\alpha_i    \\
 \cos\phi_i \cos\alpha_i+ \cos\theta_i \sin\phi_i \sin\alpha_i \\
 -\sin\theta_i \sin\alpha_i
\end{array}\right),
\end{equation}
where $\alpha_i$ is uniformly distributed in $[0,2\pi]$. 

The sum in Eq.~(\ref{Bsim}) extends over $N$ values of the wave number $k=||\mathbf{k}||$. 
The wave numbers $k_i$ are separated in such a way that $\log(k_{i+1}/k_i)=\Delta$, where 
$\Delta$ is a constant \cite{Giacalone:94}. $N=1000$ is used in these simulations.

The amplitude $A(k)$ is determined by the type of turbulence. Following Ref.~\cite{Harari:02},
it is given by
\begin{eqnarray}
\label{Ak}
A^2(k) &=& B^2_{rms} \frac{(n-1) (2\pi/L_{max})^{n-1}}{1-(L_{min}/L_{max})^{n-1}}\ \delta k  \nonumber \\
&& \times \left\{
\begin{array}{ll}
 k^{-n}  & \ \ \ k \in [2\pi/L_{max},2\pi/L_{min}], \\
 0       & \ \ \ \textrm{otherwise},
\end{array} \right.
\end{eqnarray}
where $n=5/3$ corresponds to the Kolmogorov spectrum, $L_{min}$ and $L_{max}$ correspond to 
the minimum and maximum wave lengths, respectively, $\delta k$ is the length of the interval
between consecutive values of $k$ defined by the discretization of the $k$ space 
[$\delta k_i=(10^\Delta-1)\ k_i$], and $B^2_{rms}=\langle ||\mathbf{B}(\mathbf{x})||^2 \rangle$.     

\begin{figure}[!th]
\centering 
\includegraphics[width=.47\textwidth]{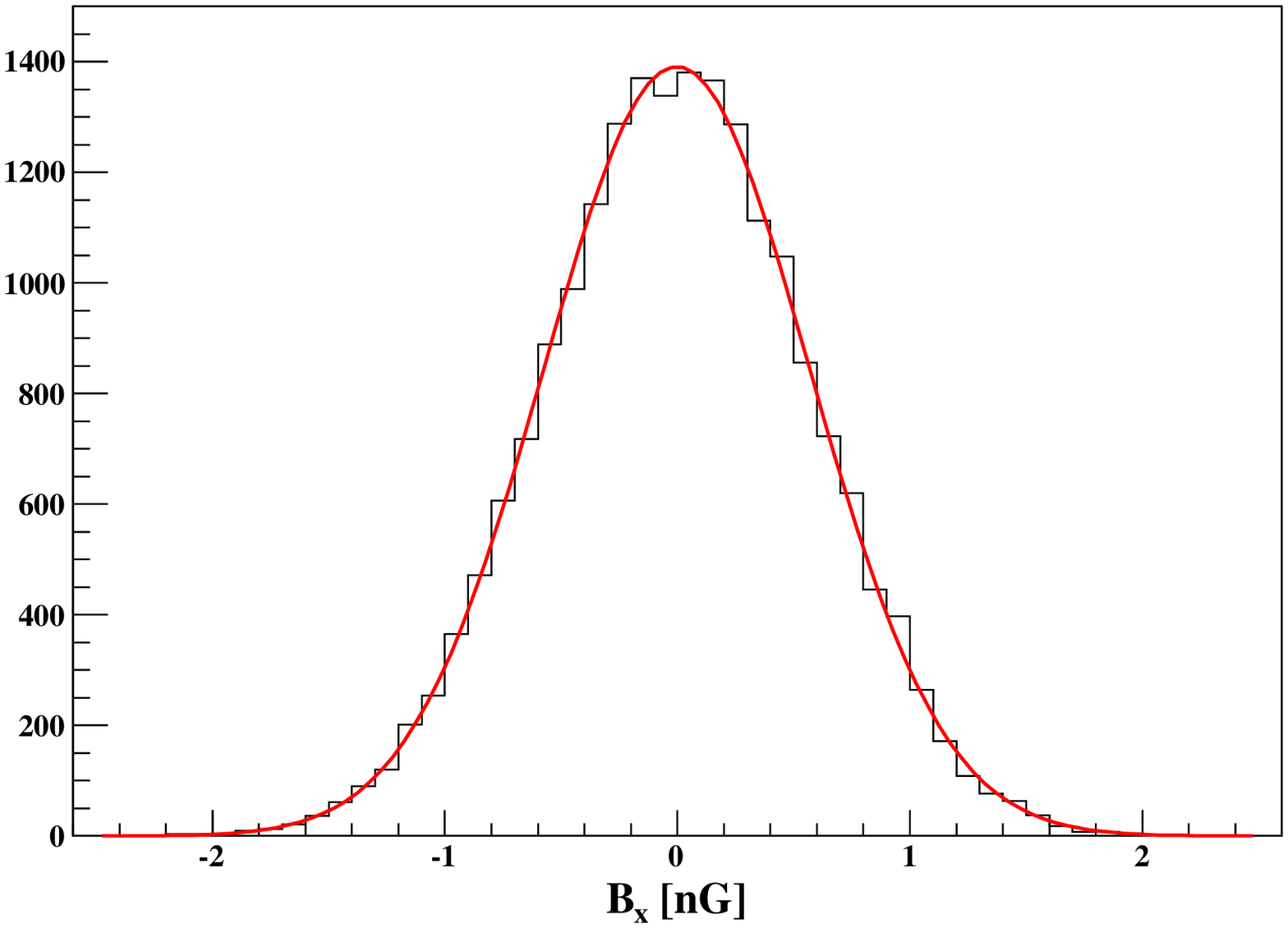}
\includegraphics[width=.47\textwidth]{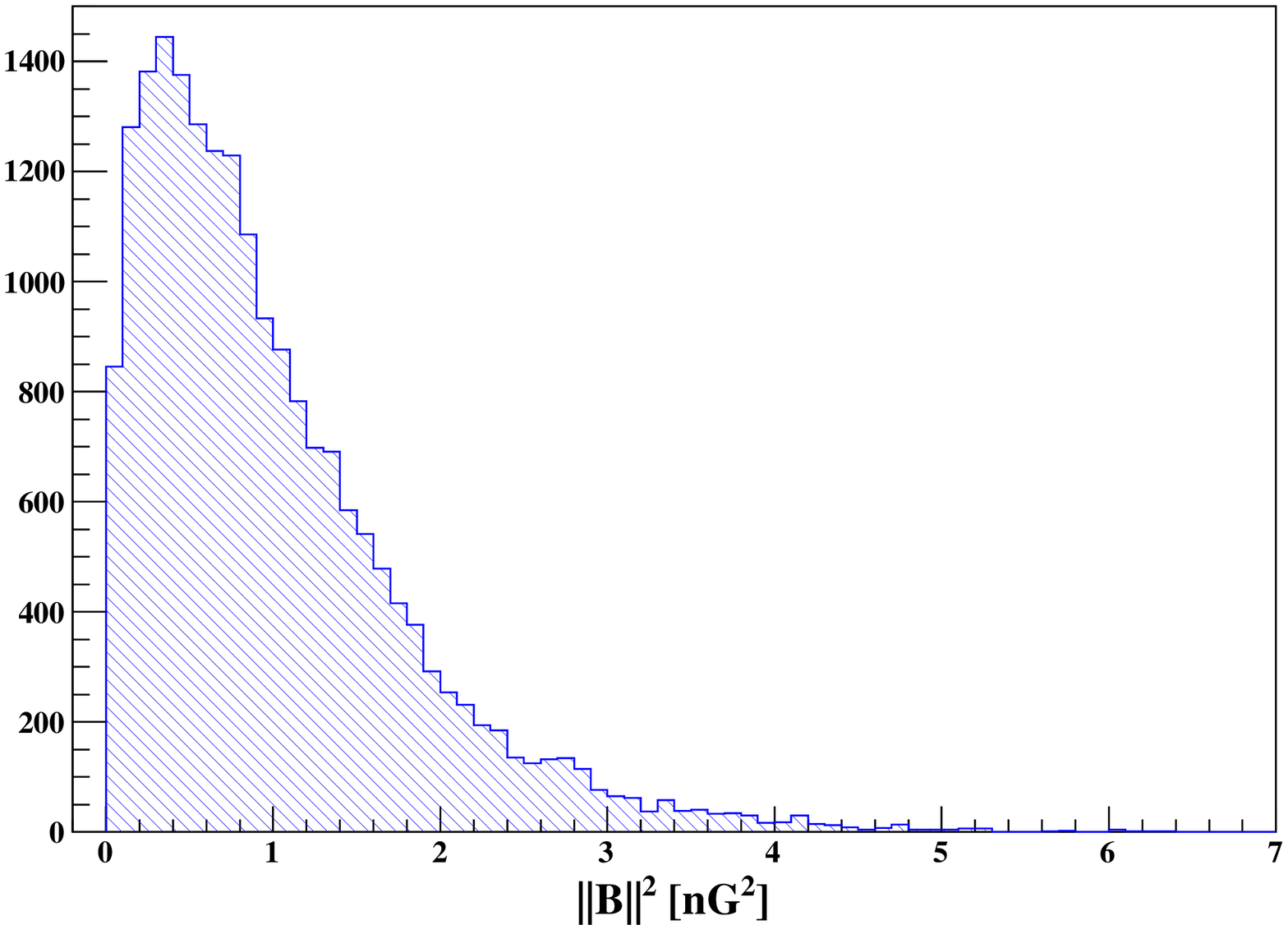}
\caption{\label{Bfield} Top panel: Distribution of the magnetic field $x$ component at $\mathbf{x}=\mathbf{0}$.
A Gaussian fit to the distribution is also shown. Bottom panel: Distribution of magnetic field squared intensity 
at $\mathbf{x}=\mathbf{0}$. Each realization of the magnetic field is such that $B_{rms} = 1$ nG and 
$L_{c} \cong 0.2$ Mpc.}
\end{figure}

Figure \ref{Bfield} shows the distribution of the $x$ component (top panel) and the squared 
intensity (bottom panel) of the magnetic field at $\mathbf{x}=\mathbf{0}$. A sample of $2\times10^5$ 
realizations of the magnetic field was used for the calculation. The parameters used are: 
$L_{max}=1$ Mpc, $L_{min}=10^{-3}$ Mpc, and $B_{rms} = 1$ nG. The coherence length of the field
is given by \cite{Harari:02},
\begin{equation}
L_{c}=\frac{(n-1)\ L_{max}}{2\ n} \times \frac{1-(L_{min}/L_{max})^n}{1-(L_{min}/L_{max})^{n-1}},
\end{equation}
which for this choice of the parameters, in which $L_{max} \gg L_{min}$, it takes the value 
$L_{c} \cong L_{max}/5 = 0.2$ Mpc. Note that IGMFs similar to this are commonly used in the literature 
\cite{Globus:07,Harari:14,Allard:14}. In the plot on the top panel of the figure a Gaussian fit to the 
distribution is also shown. Although not shown in the figure, the $y$ and $z$ components of the magnetic 
field also follow a Gaussian distribution. As expected, the mean value corresponding to the distribution 
on the bottom panel of the figure is $\sim 1$ nG.     

\begin{figure}[!th]
\centering 
\includegraphics[width=.47\textwidth]{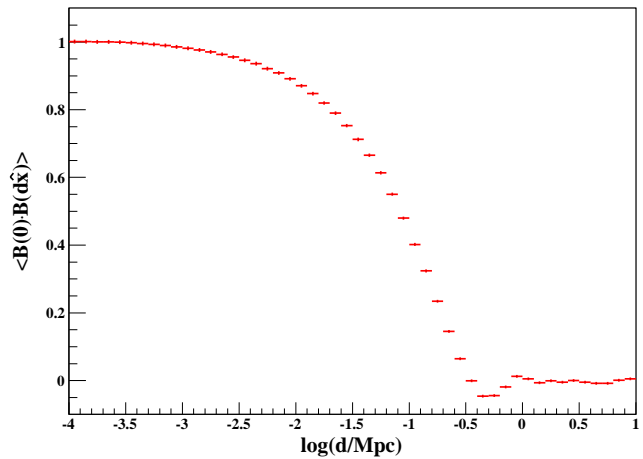}
\caption{\label{BLcoh} Mean value of $\mathbf{B}(\mathbf{0})\cdot \mathbf{B}(d \cdot \mathbf{\hat{x}})$ as a function 
of the logarithm of the distance to the origin of coordinates $d$. Each realization of the magnetic field is such 
that $B_{rms} = 1$ nG and $L_{c} \cong 0.2$ Mpc.}
\end{figure}

Figure \ref{BLcoh} shows the mean value of the scalar product between the magnetic field evaluated at 
$\mathbf{x}=\mathbf{0}$ and at $\mathbf{x}=d \cdot \mathbf{\hat{x}}$ ($\mathbf{\hat{x}}$ is a unit vector pointing 
to the $x$-axis direction) as a function of the logarithm of the distance to the coordinates origin, $d$. It can be
seen from the figure that the average of the scalar product goes to zero for distances of the order of or larger
than the coherence length of the magnetic field. 

As mentioned before, different spatial distributions of cosmic ray sources produce different energy spectra at Earth. 
Therefore, the energy spectra of a single source as a function of its distance to Earth is required in order to calculate 
the energy spectrum for a given spatial distribution of sources. For that purpose, CRPropa (version 2.0.4) \cite{CRPropa}
is used. CRPropa is a program written to propagate nuclei, nucleons, photons, and neutrinos in the intergalactic medium. 
A complete set of processes undergone by the particles propagating in the intergalactic medium are implemented in that 
program. The most important processes are photo-pion production (the SOPHIA~\cite{bib:sophia} code, available in CRPropa, 
is used), pair production, and photodisintegration in both the cosmic microwave background and the extragalactic background 
light. In particular, it is possible to propagate particles in the presence of an intergalactic magnetic field which can be 
provided externally. In this calculation, a 10 Mpc side cubic simulation box is used to inject the cosmic ray particles. The 
simulation box is filled with a turbulent magnetic field calculated by following the method described above. The 3D option is 
used for the simulation in which the trajectories of charged particles that deviate in the magnetic field are calculated 
numerically by solving the corresponding differential equations (see Ref.~\cite{CRPropa} for details). Periodic boundary 
conditions are used for the simulations. The particles are uniformly distributed in the whole volume and propagated thereafter. 
They are followed for a time of 4 Gpc/$c$, where $c$ is the speed of light. The particles are recorded every time they enter 
to a sphere of 1 Mpc around the observer (see Ref.~\cite{CRPropa} for details on the detection algorithm).  

The propagation of protons and iron nuclei is simulated for three cases: a null magnetic field, $B_{rms} = 5$ nG, and 
$B_{rms} = 10$ nG. The differential injection spectra for both types of nuclei considered follows a power law with spectral 
index $\gamma = 1$ ($\phi\propto E^{-\gamma}$). The energy range extends from $10^{19}$ to $10^{22}$ eV. Note that it 
is possible to obtain a different injection spectrum by appropriately weighting each particle that reaches the observer.
For instance, a power law with spectral index $\gamma=-2$ combined with an exponential cutoff is obtained by weighting each 
particle with $w_i=E_{0,i}^{-1}\times\exp(-E_{0,i}/E_{cut})$, where $E_{0,i}$ is the energy of the parent particle injected 
at the source corresponding to $i$th particle observed at Earth and $E_{cut}$ is the cutoff energy. The output of these 
simulations is used to calculate the energy spectrum as a function of source distance.

Note that the 3D propagation in a non-null intergalactic magnetic field implemented in CRPropa does not include the effects 
of the expansion of the Universe and the evolution of the photon backgrounds with redshift. However, these effects become 
important at energies smaller than a few $10^{19}$ eV \cite{Ahlers:13}. At higher energies they are negligible, because the 
flux at Earth is dominated by local sources, which is due to the interactions undergone by the cosmic rays during propagation.    

\section{Ensemble fluctuations and the intergalactic magnetic field}
\label{EF}

The injection spectrum considered is given by a power law with an exponential cutoff,
\begin{equation}
\phi(E) = C_0\ E^{-\gamma} \exp(-E/E_{cut}),
\end{equation}  
where $C_0$ is a constant, $\gamma$ is the spectral index, and $E_{cut}$ is the cutoff energy. From the simulations 
described in Sec.~\ref{Sims} it is possible to obtain the number of particles of type $A$ that reach the observer 
per unit of energy $E$. It is denoted as $K(E,A|r,E_{cut},\gamma,A_0,\mathbf{B})$, where $r$ is the injection 
distance, $A_0$ is the type of particles injected by the sources (proton or iron in this work), and $\mathbf{B}$ is 
the realization of the IGMF considered.

For a given configuration of the IGMF, the flux observed at Earth depends on the density and spatial distribution of 
the sources. Therefore, the flux corresponding to a given distribution of sources and density can be calculated as
\begin{equation}
J_{A,m}(E) = C\ \sum_{i=1}^{N_s} \frac{K(E,A|r_{m,i},E_{cut},\gamma,A_0,\mathbf{B})}{r_{m,i}^2},
\end{equation}  
where $C$ is a normalization constant, $N_s$ is the number of sources in the volume under consideration, and $r_{m,i}$
is the distance of the $i$th source of the sample $m$ ($m=1,...,M$, where $M$ is the number of samples considered). 
Here a constant density of sources is considered. Therefore, $N_s$ is sampled from a Poisson distribution with mean value
\begin{equation} 
\mu=\frac{4\pi}{3} n_s (R^3-r_{min}^3).   
\end{equation}  
The distance of a given source $r$ is sampled form
\begin{equation} 
P(r)= \frac{3}{R^3-r_{min}^3} \times
\left\{
\begin{array}{ll}
 r^2     & \ \ \ r \in [r_{min},R], \\
 0       & \ \ \ \textrm{otherwise}.
\end{array} \right.
\end{equation}  
Note that $r_{min}$ is the minimum distance allowed for a source \cite{Ahlers:13} and 
$R$ is taken as 4000 Mpc.

The density of UHECR sources is unknown. A lower limit on the local density of sources has been obtained 
from the analysis of the data taken by The Pierre Auger Observatory \cite{Auger:13}. Assuming that the 
sources are distributed uniformly and by using the two-point autocorrelation function, a lower limit of 
$\sim (0.06-5) \times 10^{-4}$ Mpc$^{-3}$ at 95 \% confidence level (C.L.) was obtained. If the sources are 
distributed by following the local matter distribution, the lower limit is $\sim (0.2-7) \times 10^{-4}$ Mpc$^{-3}$
at 95 \% C.L. Therefore, motivated by these results the values of the density of sources $n_s=10^{-5}$ Mpc$^{-3}$ 
and $n_s=10^{-4}$ Mpc$^{-3}$ are considered in this work.

\begin{figure}[!th]
\centering 
\includegraphics[width=.47\textwidth]{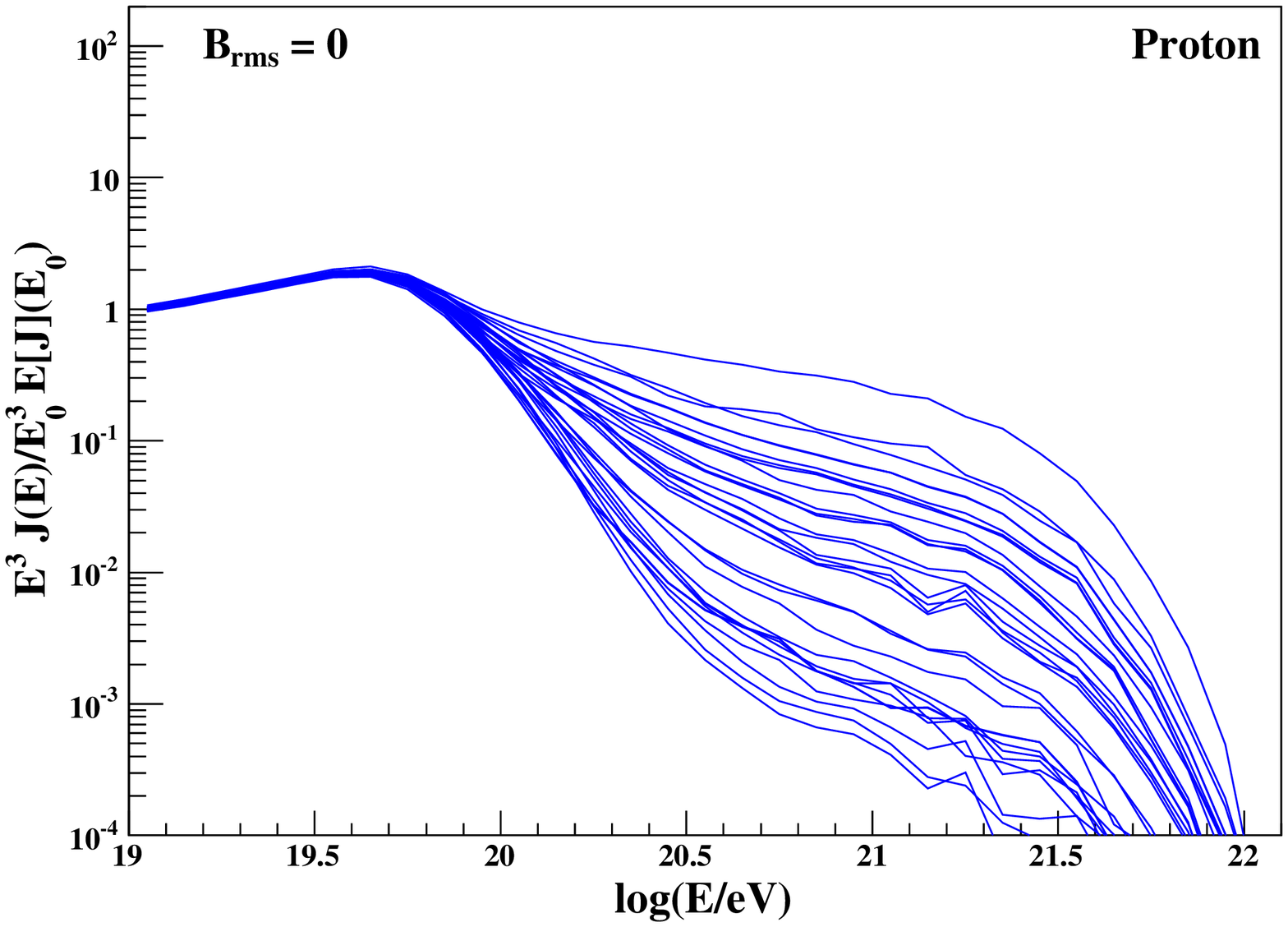}
\includegraphics[width=.47\textwidth]{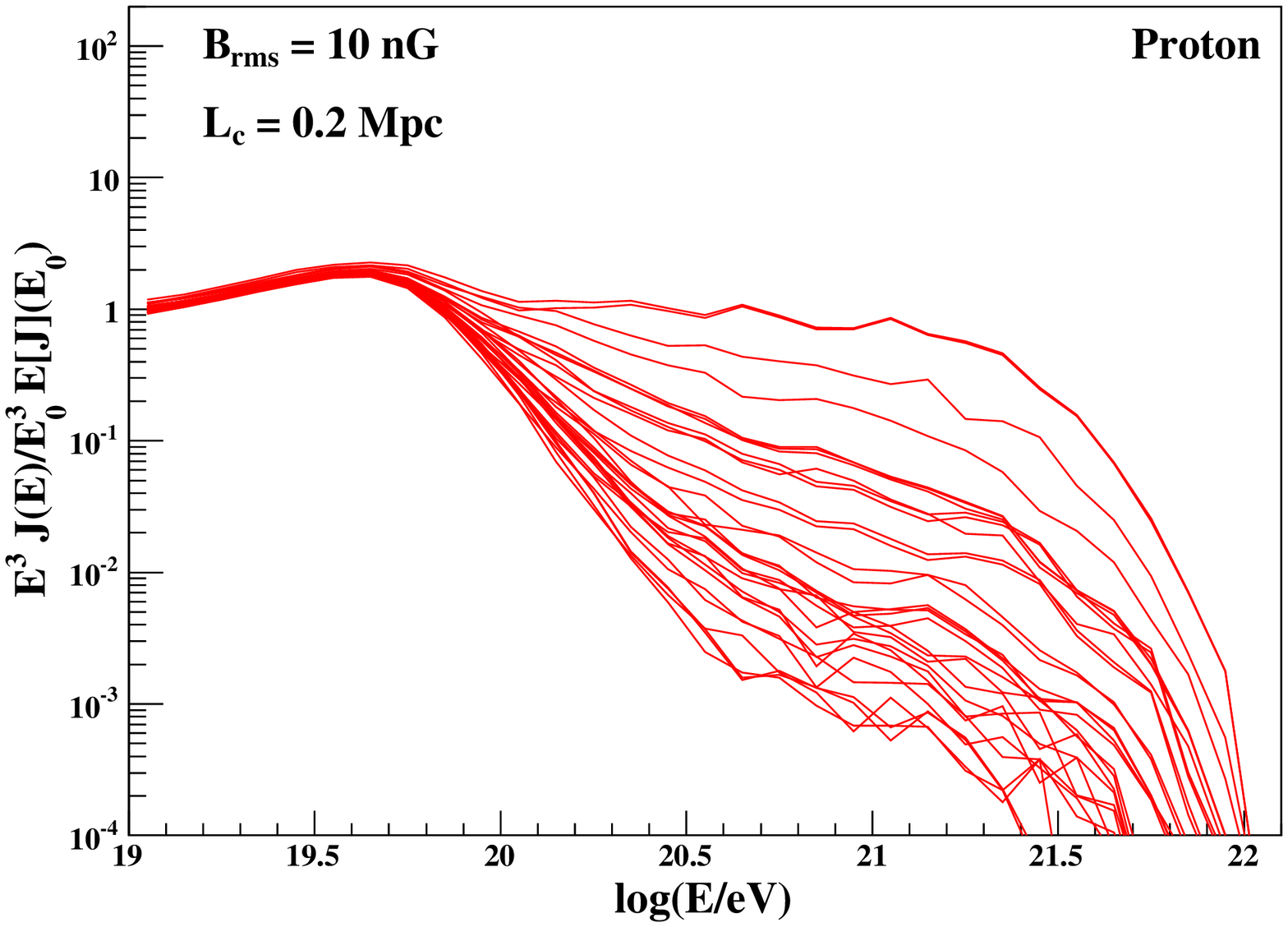}
\caption{\label{SPr} Cosmic ray energy spectra multiplied by $E^3$ as a function of the logarithm of energy 
for $B_{rms}=0$ (top panel) and $B_{rms}=10$ nG (bottom panel). The sources inject only protons into the 
intergalactic medium. The spectra are normalized to their mean value multiplied by $E^3$ evaluated at 
$E_0 = 10^{19}$ eV. The density of sources is $n_s=10^{-5}$ Mpc$^{-3}$, $E_{cut}=10^{21}$ eV, $\gamma=2.2$, and 
$r_{min}=3$ Mpc.}
\end{figure}

Figure \ref{SPr} shows the cosmic ray energy spectra corresponding to different realizations of the distribution
of sources. It is assumed in this case that the sources inject only protons into the intergalactic medium. 
Two different configurations of the IGMF, $B_{rms}=0$ (top panel) and $B_{rms}=10$ nG (bottom panel), are shown. 
The density of sources used for the calculation is $n_s=10^{-5}$ Mpc$^{-3}$ and $r_{min}=3$ Mpc. Following 
Ref.~\cite{Ahlers:13}, we take $E_{cut}=10^{21}$ eV and $\gamma=2.2$.   
From the figure, it can be seen that in both cases the spectra present a strong suppression at $E \sim 10^{19.6}$ eV,
which is of the order of the energy threshold of photopion production of protons interacting with low energy photons
of the cosmic microwave background. It can also be seen that the ensemble fluctuations of the spectrum increase with 
energy becoming quite large beyond the beginning of the suppression. This is because, for energies larger than the 
photopion production threshold, only nearby sources can contribute to the total flux. In this case, there is no evident 
influence of a non-null IGMF on the energy spectrum. 

From a sample of energy spectra, it is possible to estimate the mean value $E[J](E)$ and the relative standard
deviation
\begin{equation} 
\varepsilon[J](E)=\frac{\sigma[J](E)}{E[J](E)},   
\end{equation}  
where $\sigma[J](E)$ is the standard deviation of the spectrum. 

\begin{figure}[!th]
\centering 
\includegraphics[width=.47\textwidth]{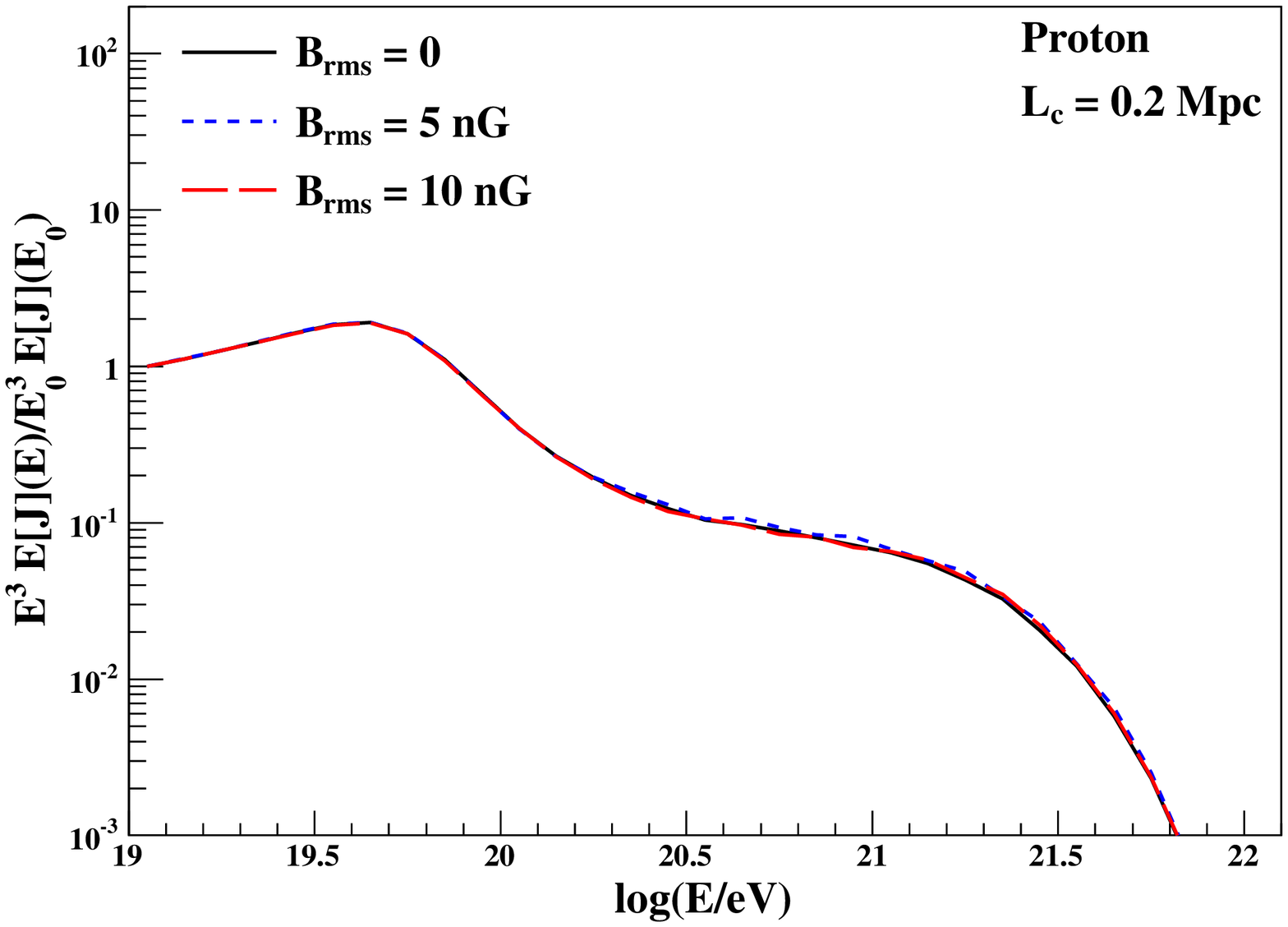}
\includegraphics[width=.47\textwidth]{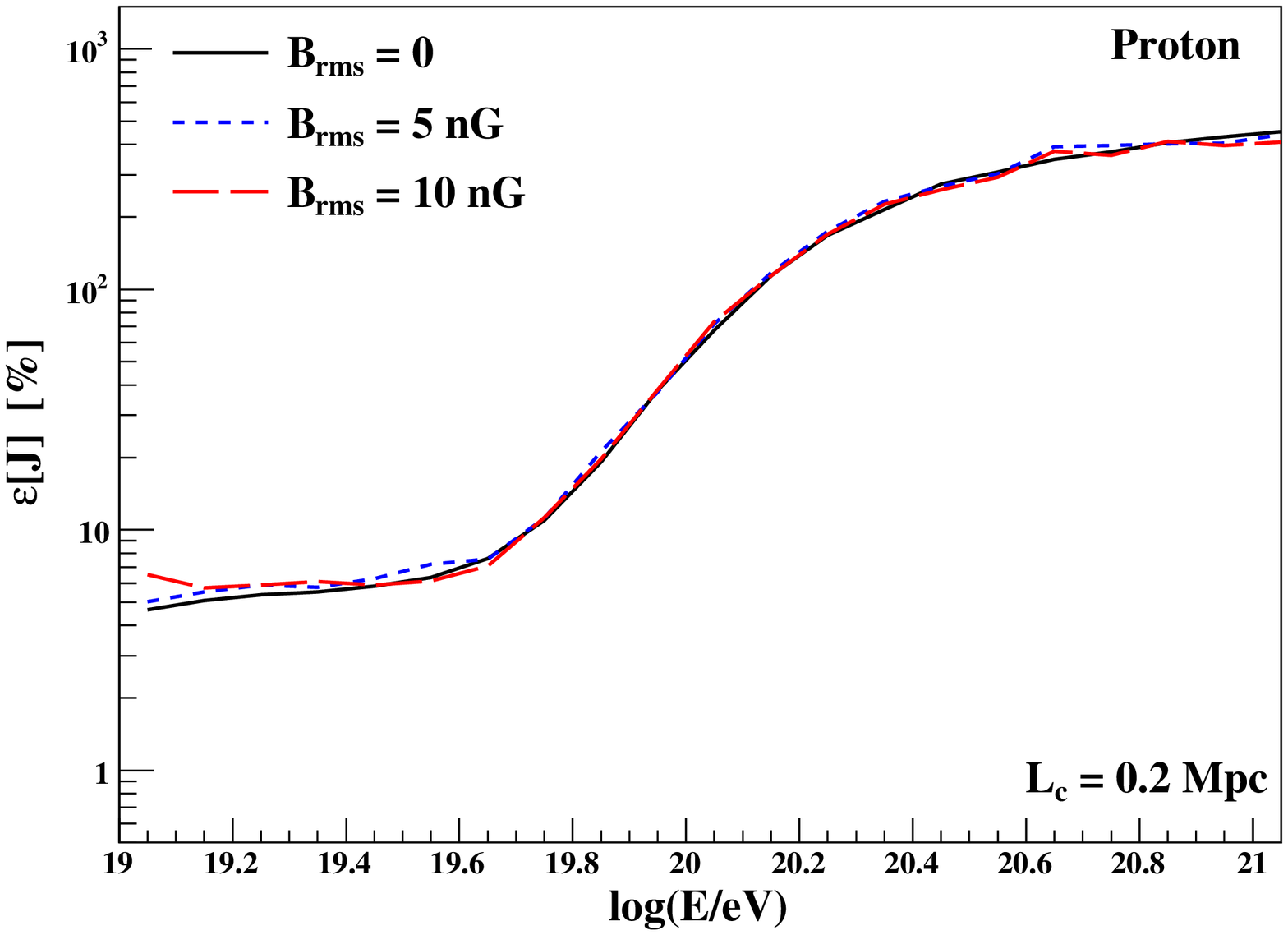}
\caption{\label{MEpsPr}Top panel: Mean value of the energy spectrum multiplied by $E^3$ as a 
function of the logarithm of energy. Bottom panel: Relative standard deviation as a function of 
the logarithm of energy. The sources inject only protons and the IGMF is such that $B_{rms}=0$, 
$5$, and $10$ nG. The density of sources is $n_s=10^{-5}$ Mpc$^{-3}$, $E_{cut}=10^{21}$ eV, 
$\gamma=2.2$, and $r_{min}=3$ Mpc.}
\end{figure}

Figure \ref{MEpsPr} shows the mean value of the energy spectra (top panel) and the relative 
standard deviation (bottom panel) as a function of the logarithm of energy. Three cases are 
considered: $B_{rms}=0$, $5$, and $10$ nG. From the figure, it can be seen that the spectra 
and the relative standard deviations are practically not affected by the IGMF. Just a small 
increase on the relative standard deviation corresponding to $B_{rms}=10$ nG can be observed 
in the energy range from $10^{19}$ to $10^{19.4}$ eV. A comparison between the results 
obtained in this work and in Refs.~\cite{Ahlers:13,AlhersICRC:13}, for the case of a null IGMF
and for sources injecting only protons, is given in Appendix \ref{App}.

\begin{figure}[!th]
\centering 
\includegraphics[width=.47\textwidth]{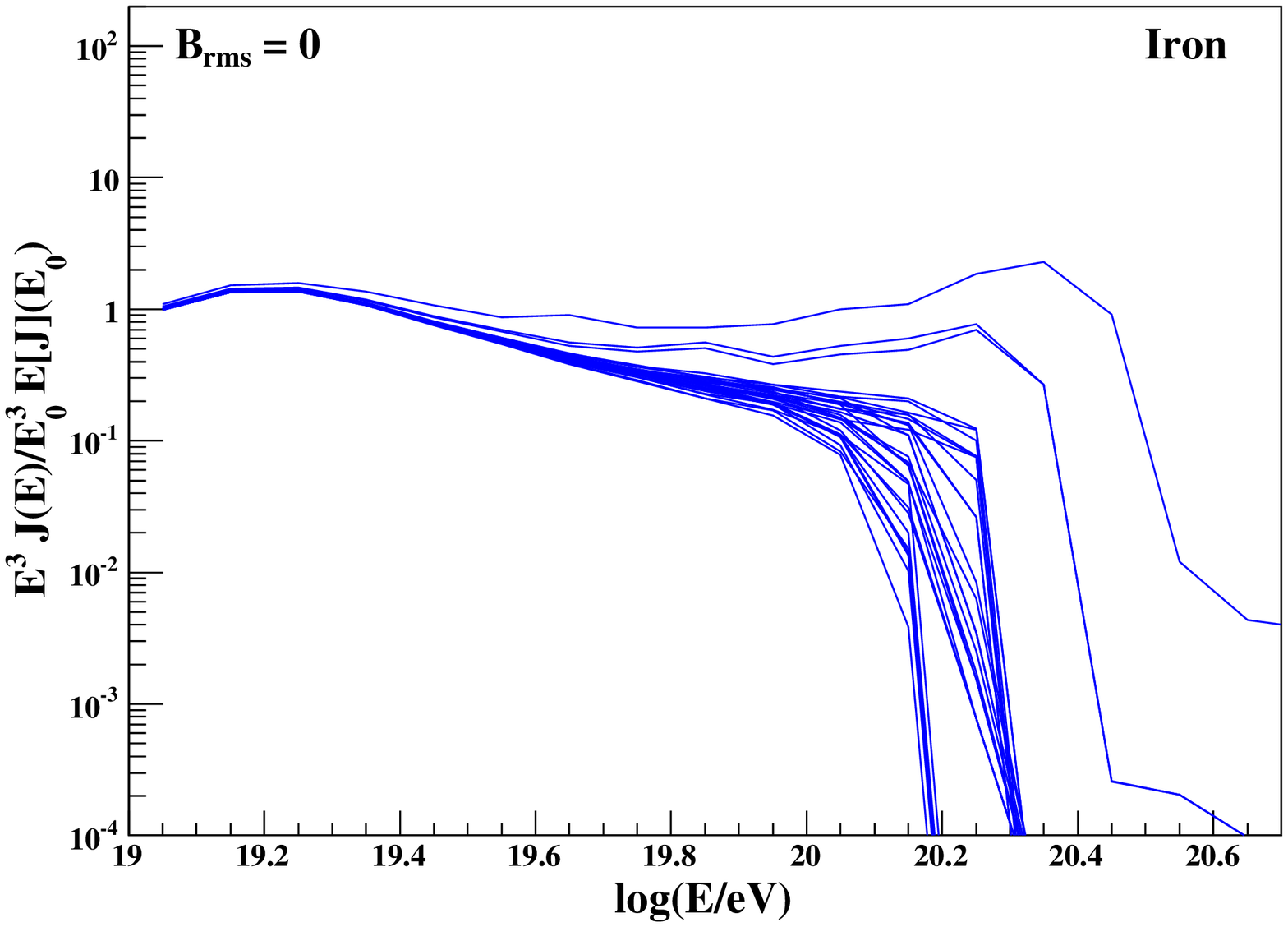}
\includegraphics[width=.47\textwidth]{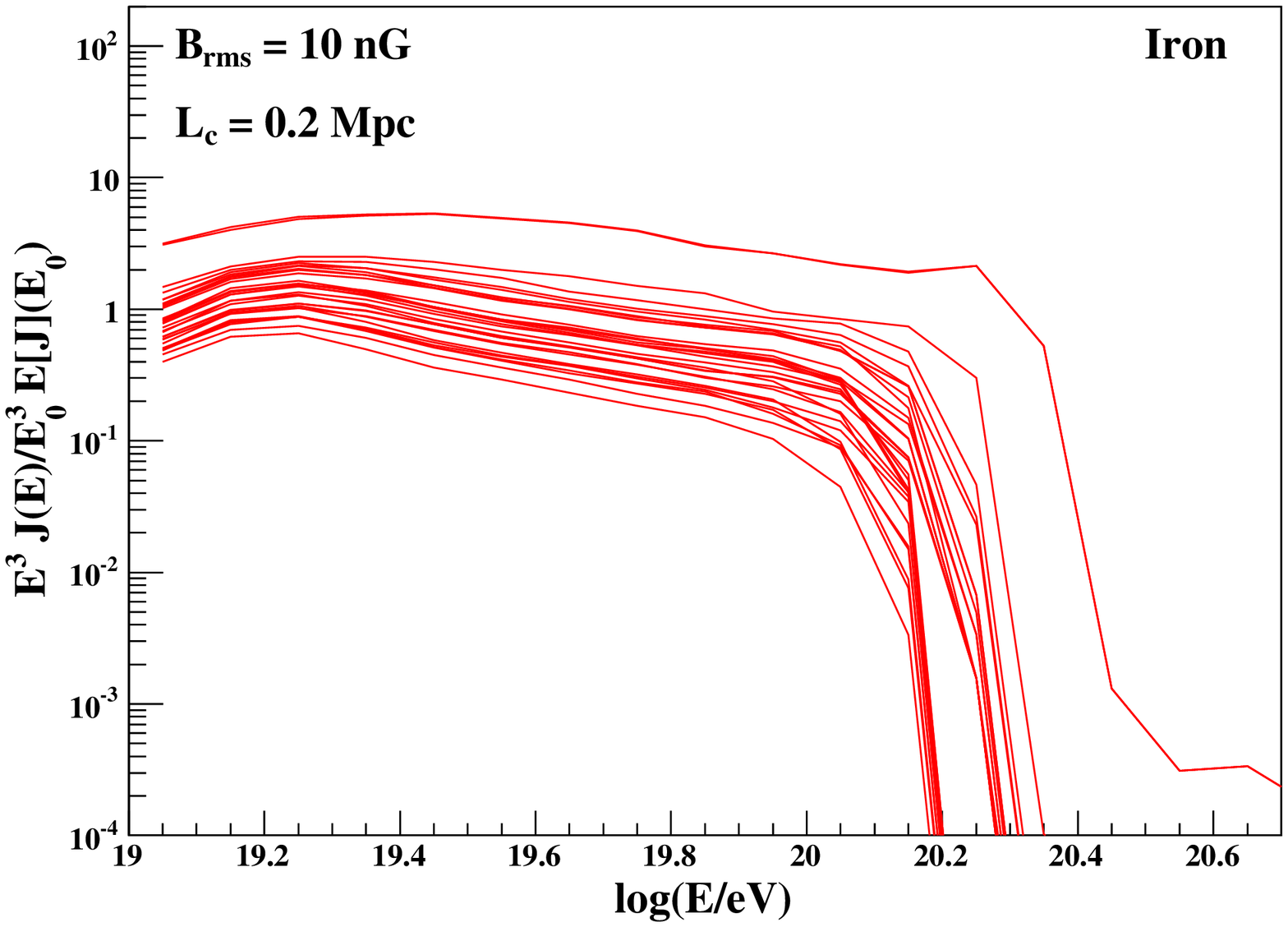}
\caption{\label{SFe} Cosmic ray energy spectra multiplied by $E^3$ as a function of the logarithm 
of energy for $B_{rms}=0$ (top panel) and $B_{rms}=10$ nG (bottom panel). The sources inject only 
iron nuclei into the intergalactic medium. The spectra are normalized to the mean value multiplied 
by $E^3$ evaluated at $E_0 = 10^{19}$ eV. The density of sources is $n_s=10^{-5}$ Mpc$^{-3}$, 
$E_{cut}=10^{21}$ eV, $\gamma=2$, and $r_{min}=3$ Mpc.}
\end{figure}

Figure \ref{SFe} shows the cosmic ray energy spectra corresponding to different realizations of the 
distribution of sources, but in this case the sources inject only iron nuclei into the intergalactic 
medium. Two different configuration of the intergalactic magnetic field are shown: $B_{rms}=0$ 
(top panel) and $B_{rms}=10$ nG (bottom panel). The density of sources used for the calculation 
is $n_s=10^{-5}$ Mpc$^{-3}$ and $r_{min}=3$ Mpc. In this case we take $E_{cut}=10^{21}$ eV and 
$\gamma=2$ \cite{Ahlers:13}. Note that it is evident that for the $B_{rms}=10$ nG case the 
ensemble fluctuations are much larger than for $B_{rms}=0$. 

\begin{figure}[!th]
\centering 
\includegraphics[width=.47\textwidth]{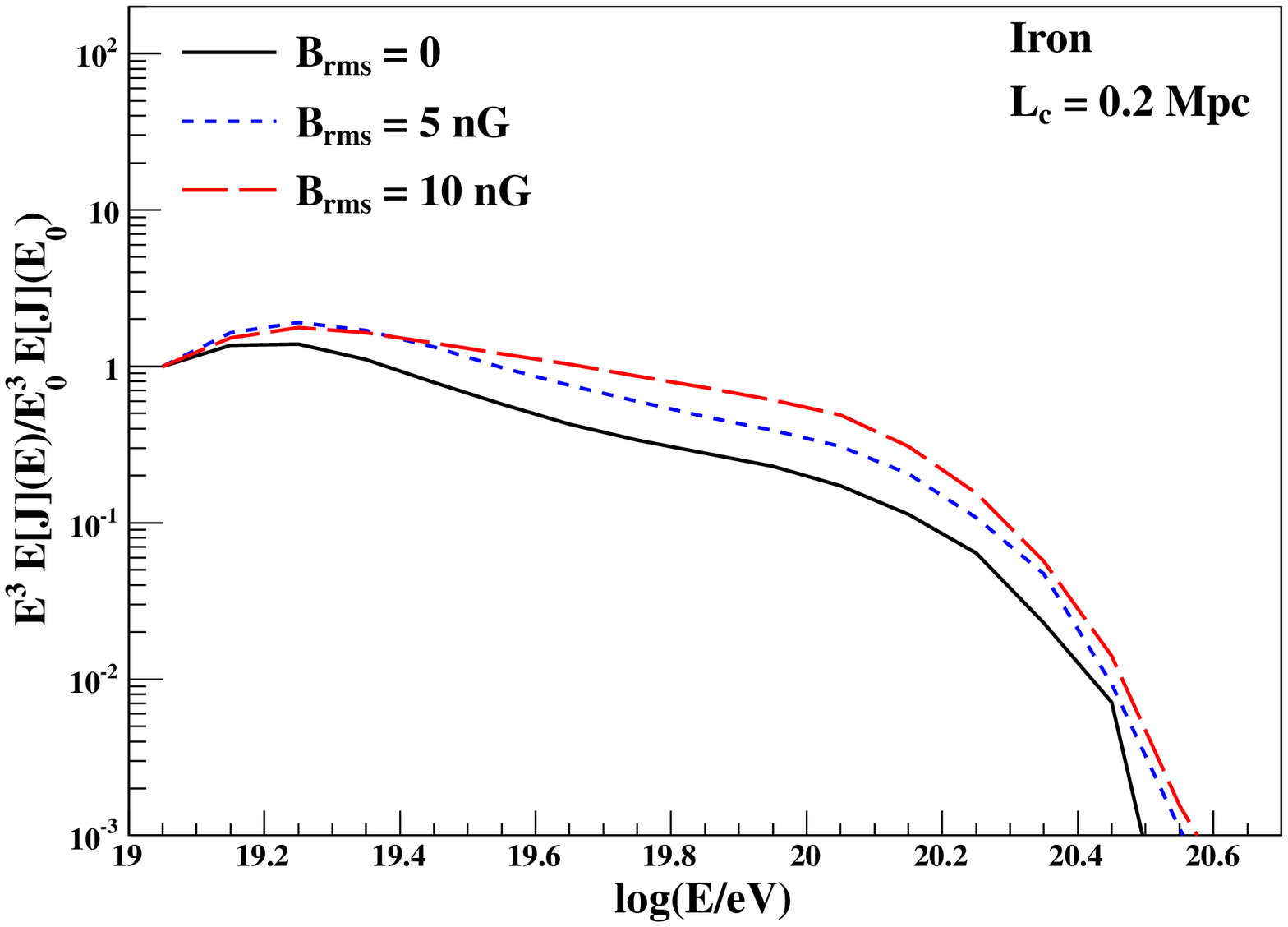}
\includegraphics[width=.47\textwidth]{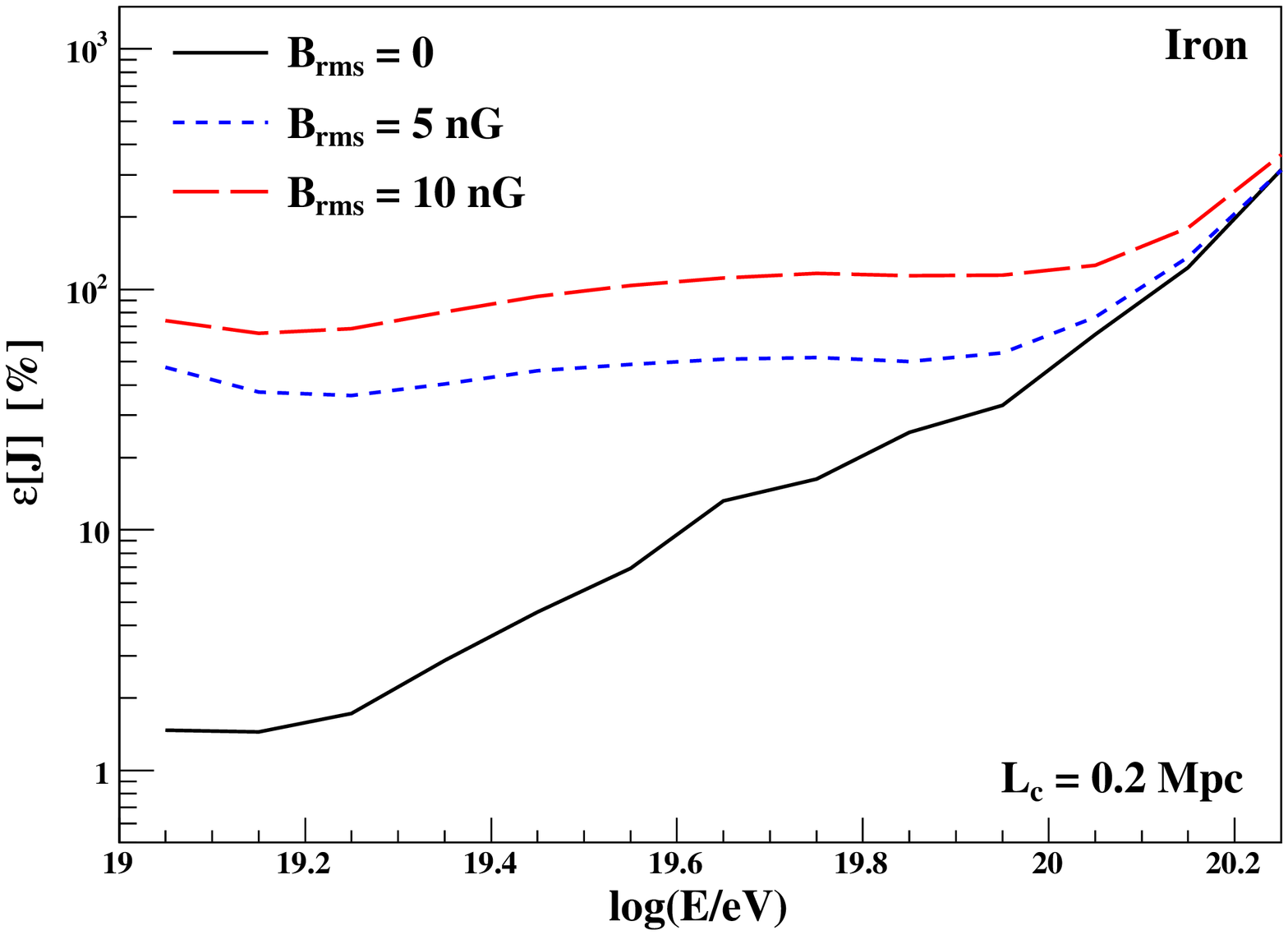}
\caption{\label{MEpsFe} Top panel: Mean value of the energy spectrum multiplied by $E^3$ as a function 
of the logarithm of energy. Bottom panel: Relative standard deviation as a function of the logarithm of 
energy. The sources inject only iron nuclei and the intergalactic magnetic field is such that $B_{rms}=0$, 
$5$, and $10$ nG. The density of sources is $n_s=10^{-5}$ Mpc$^{-3}$, $E_{cut}=10^{21}$ eV, $\gamma=2$, and 
$r_{min}=3$ Mpc.}
\end{figure}

Figure \ref{MEpsFe} shows the mean value of the energy spectra (top panel) and the relative 
standard deviation (bottom panel) as a function of the logarithm of energy for the case in which 
the sources inject only iron nuclei into the intergalactic medium. In this case the presence of 
a non-null IGMF modifies the shape of the mean energy spectrum. Also the ensemble fluctuations 
increase significantly at low energy. From the figure, it can be seen that at $E \sim 10^{19}$ eV the 
relative standard deviation for $B_{rms} \geq 5$ nG is more than one order of magnitude larger 
than for $B_{rms}=0$. The relative standard deviation curves for non-null IGMF get closer to the 
curve corresponding to a null IGMF when energy increases. That is in such a way that for 
$E\gtrsim 10^{20}$ eV the differences are small. A comparison between the results obtained in this 
work and in Refs.~\cite{Ahlers:13,AlhersICRC:13}, for the case of a null IGMF and for sources injecting 
only iron nuclei, is given in Appendix \ref{App}.  

\begin{figure}[!th]
\centering 
\includegraphics[width=.47\textwidth]{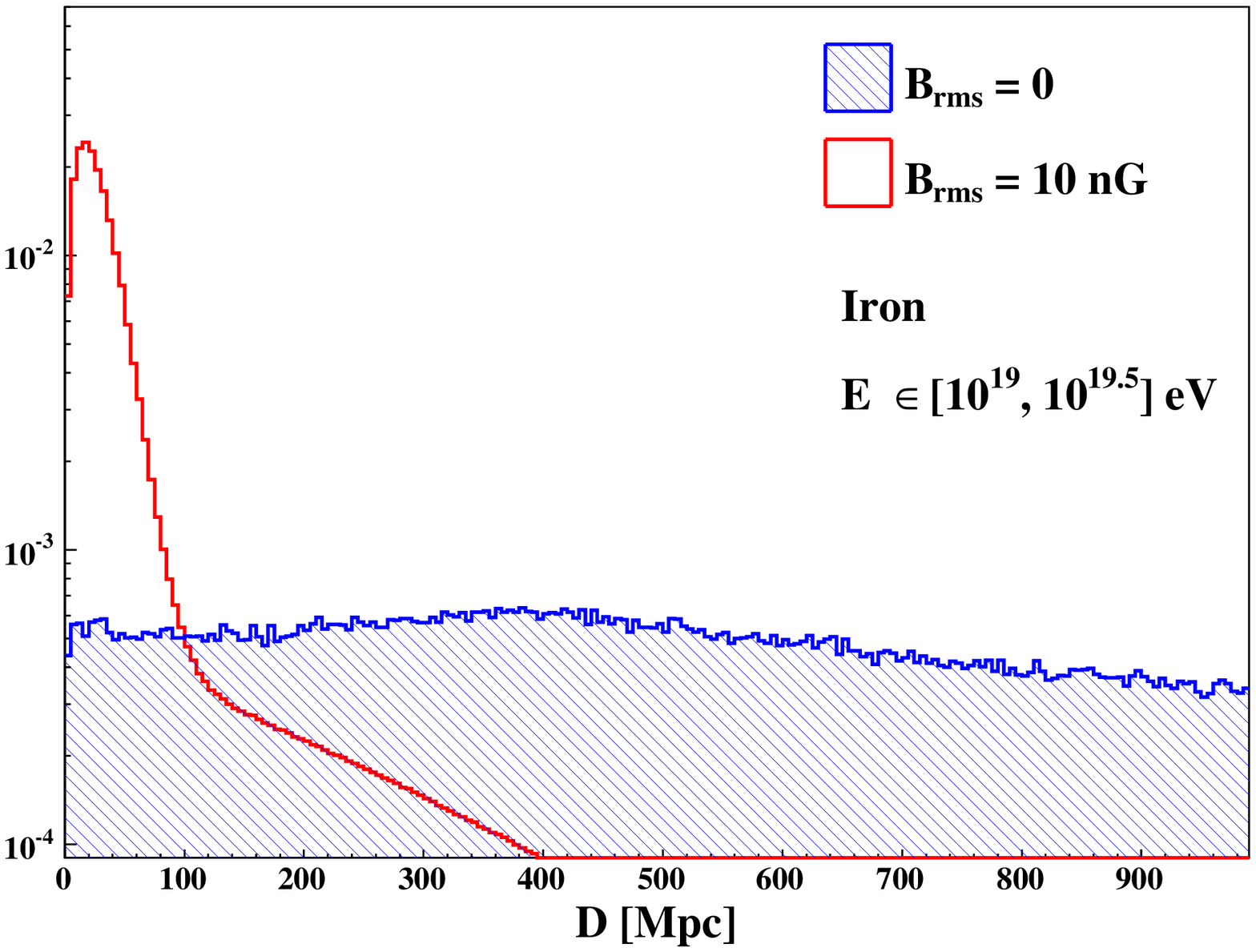}
\includegraphics[width=.47\textwidth]{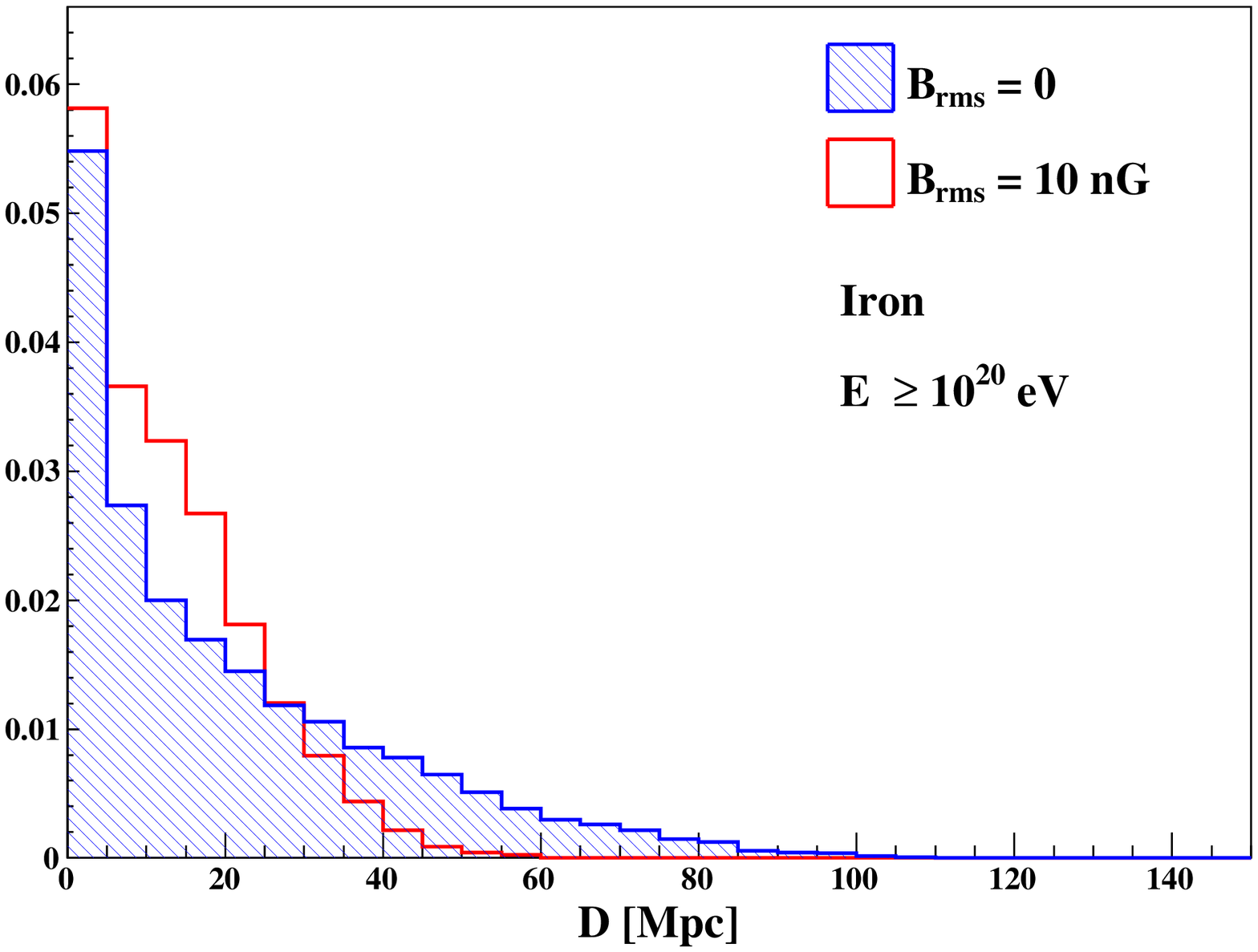}
\caption{\label{Ddist} Distribution of source distance for a null IGMF and for $B_{rms} = 10$ nG. The sources 
inject only iron nuclei. The spectral index is $\gamma=2$ and $E_{cut} = 10^{21}$ eV. Top panel: Low energy 
interval ($E \in [10^{19}, 10^{19.5}]$ eV). Bottom panel: High energy interval ($E\geq10^{20}$ eV).}
\end{figure}

The increase of the ensemble fluctuations of the spectrum at low energies can be understood from the fact 
that, in the presence of a non-null IGMF, cosmic rays travel larger distances before reaching the observer due 
to the deviations suffered by the interaction with the magnetic field. Therefore, the sources that can 
contribute to the flux observed at Earth are closer than in the case of a null IGMF. In fact, at any given 
energy, fewer sources contribute to the spectrum, increasing its ensemble fluctuations, and this effect is 
more notorious as the energy decreases. Figure \ref{Ddist} shows the distribution of source distance $D$ 
corresponding to cosmic rays that reach the observer for sources that inject only iron nuclei into the 
intergalactic medium. In the top panel of the figure, the distributions corresponding to a null IGMF and 
$B_{rms} = 10$ nG, for $E \in [10^{19}, 10^{19.5}]$ eV, are shown. It can be seen that the distribution 
for $B_{rms} = 10$ nG is strongly peaked at $\sim 40$ Mpc and then goes to zero quite fast. Meanwhile, for 
the case of a null IGMF, the distribution is quite flat, allowing distant sources to contribute to the spectrum. 
The bottom panel of the figure shows the distributions of source distance but for $E\geq10^{20}$ eV. In this 
case, both distributions have a maximum at $D=0$ and go to zero quite fast. However, the one corresponding 
to $B_{rms} = 10$ nG is more concentrated around zero, which means that also in this case the sources that 
dominate the spectrum for $B_{rms} = 10$ nG are closer than for the case of a null IGMF. Note that in this 
case the difference is smaller compared to the case corresponding to lower energies. This is due to the 
fact that charged particles of larger energies are less affected by the IGMF. For protons, the effect is 
negligible.    

For the case in which the sources inject heavy nuclei, also the composition profile at Earth is affected by the 
IGMF. The composition observed at Earth can be characterized by the average of the logarithm of the mass number 
$A$, which is given by
\begin{equation}
\langle \ln A \rangle (E) =\frac{\sum_{A} \ln A\ J_A(E)}{\sum_{A} J_A(E)}. 
\end{equation}
Note that $\langle \ln A \rangle$ also depends on the realization of the spatial distribution of sources.

\begin{figure}[!th]
\centering 
\includegraphics[width=.47\textwidth]{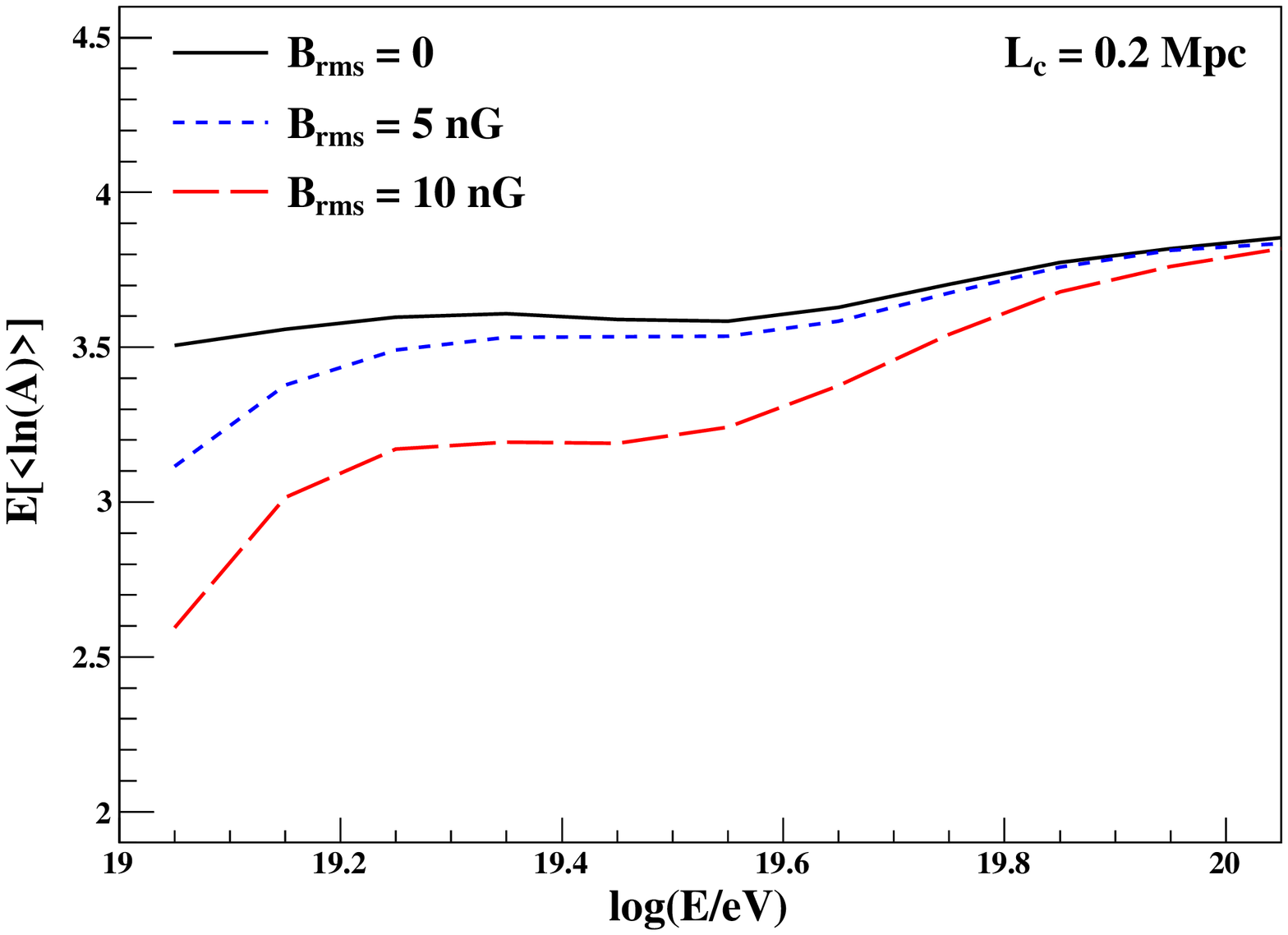}
\includegraphics[width=.47\textwidth]{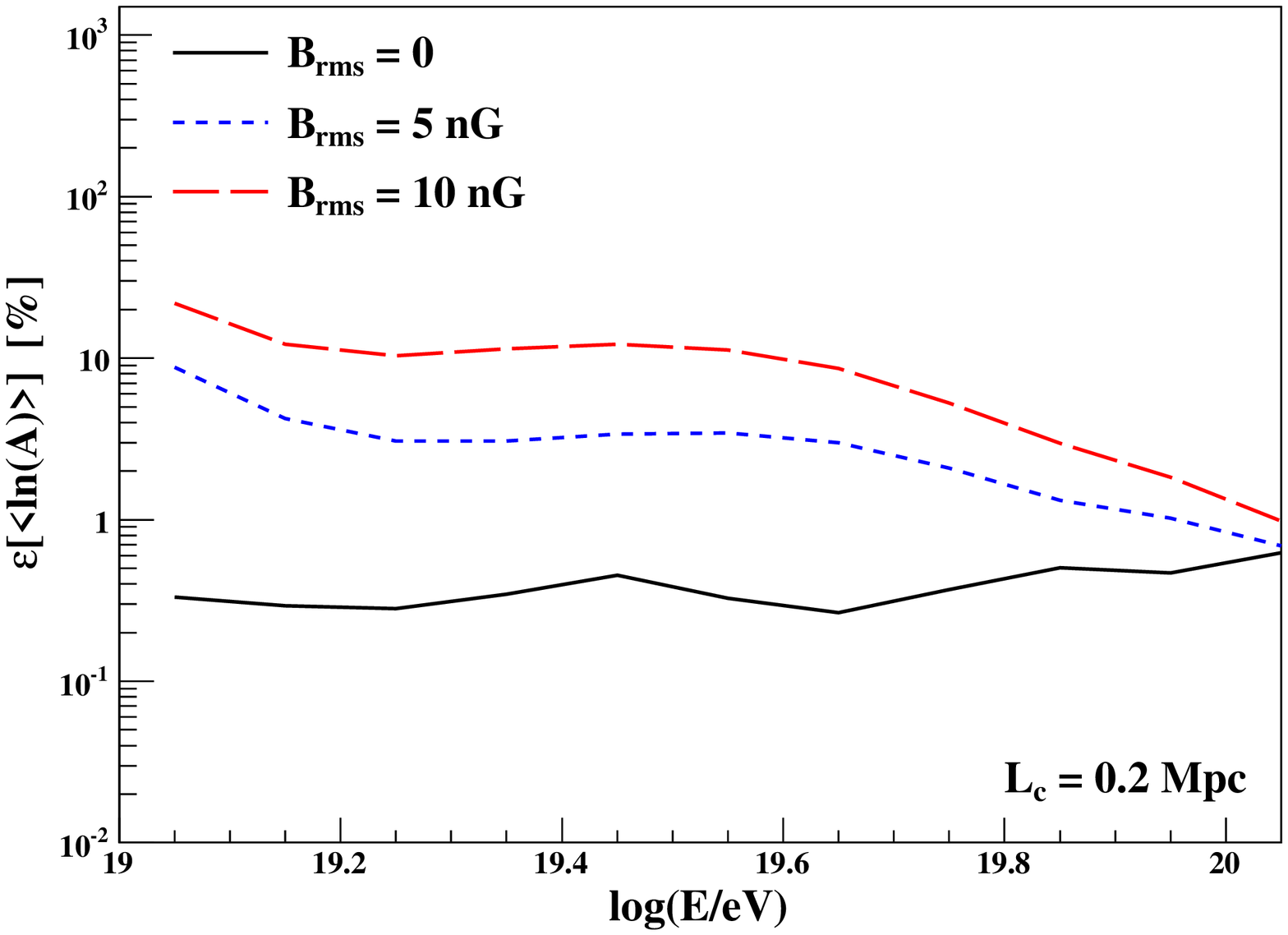}
\caption{\label{Compo} Top panel: Mean value of $\langle \ln A \rangle$ as a function of the logarithm of energy. 
Bottom panel: Relative standard deviation of $\langle \ln A \rangle$ as a function of the logarithm of energy.
The spectral index is $\gamma=2$, $E_{cut} = 10^{21}$ eV, $r_{min}=3$ Mpc, and $n_s=10^{-5}$ Mpc$^{-3}$.}
\end{figure}

As mentioned before, the cosmic rays travel larger distances in the presence of a non-null IGMF. Therefore, the 
probability to suffer a given interaction with the radiation field of the intergalactic medium is higher. In 
particular, they can suffer photodisintegration. As a result, a lighter composition is expected at Earth 
\cite{Taylor:11}. The top panel of Fig.~\ref{Compo} shows the mean value of $\langle \ln A \rangle$ as a 
function of the logarithm of energy. From this plot, it can be seen that the composition gets lighter for 
increasing values of $B_{rms}$.        

The ensemble fluctuations of the composition profile are also affected by the IGMF. The bottom panel of Fig.~\ref{Compo} 
shows the relative standard deviation of $\langle \ln A \rangle$ as a function of the logarithm of energy. As in the case 
of the energy spectrum, the ensemble fluctuations are larger for increasing values of $B_{rms}$.

Note that also, in the case of the composition profile and its relative standard deviation, the curves corresponding to 
non-null IGMFs get closer to the one corresponding to a null IGMF for increasing values of energy. As mentioned before, 
this is caused by the diminution of the magnetic field effects suffered by the charged cosmic rays for increasing values 
of energy.    

\begin{figure}[!th]
\centering 
\includegraphics[width=.47\textwidth]{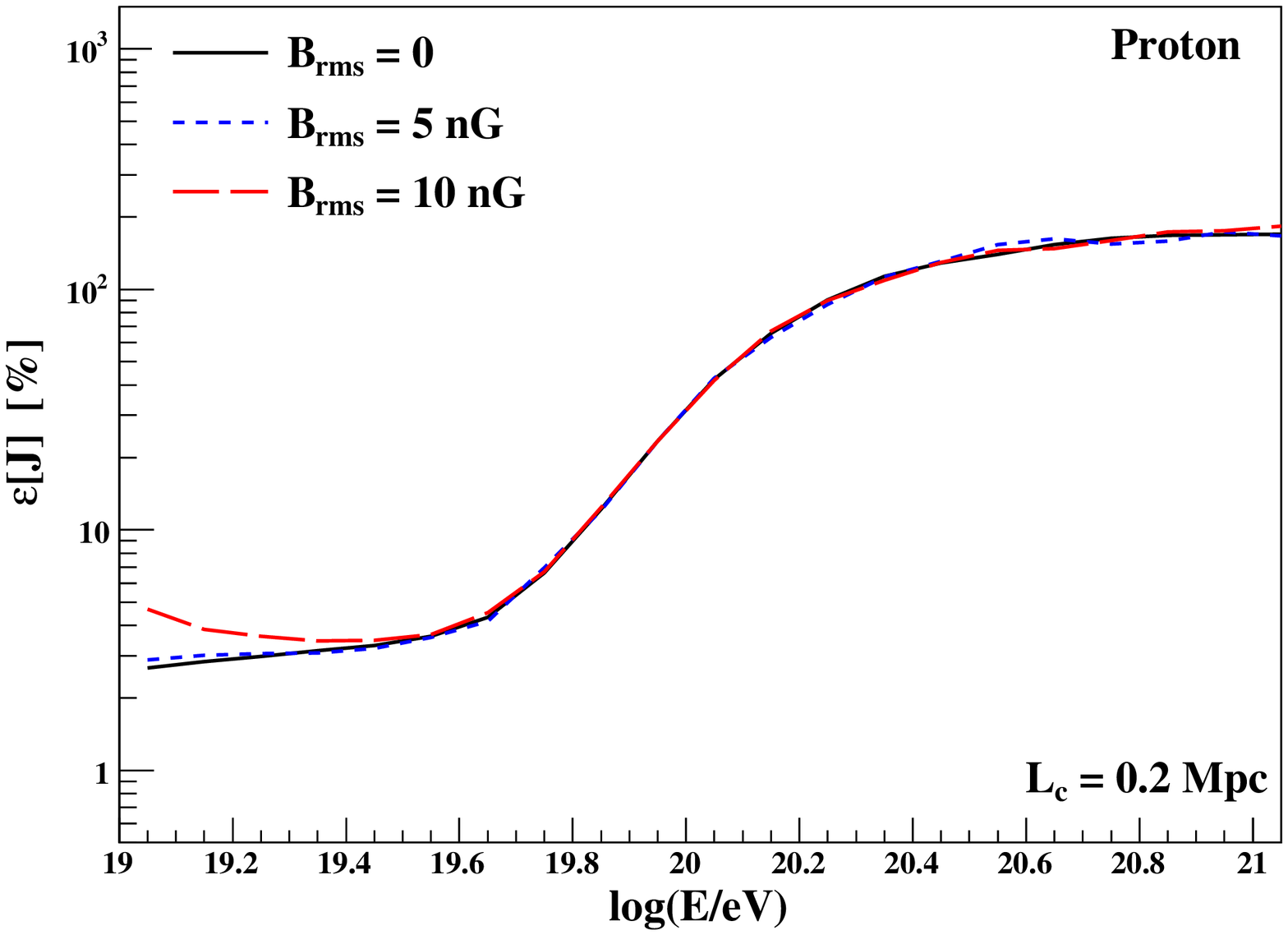}
\includegraphics[width=.47\textwidth]{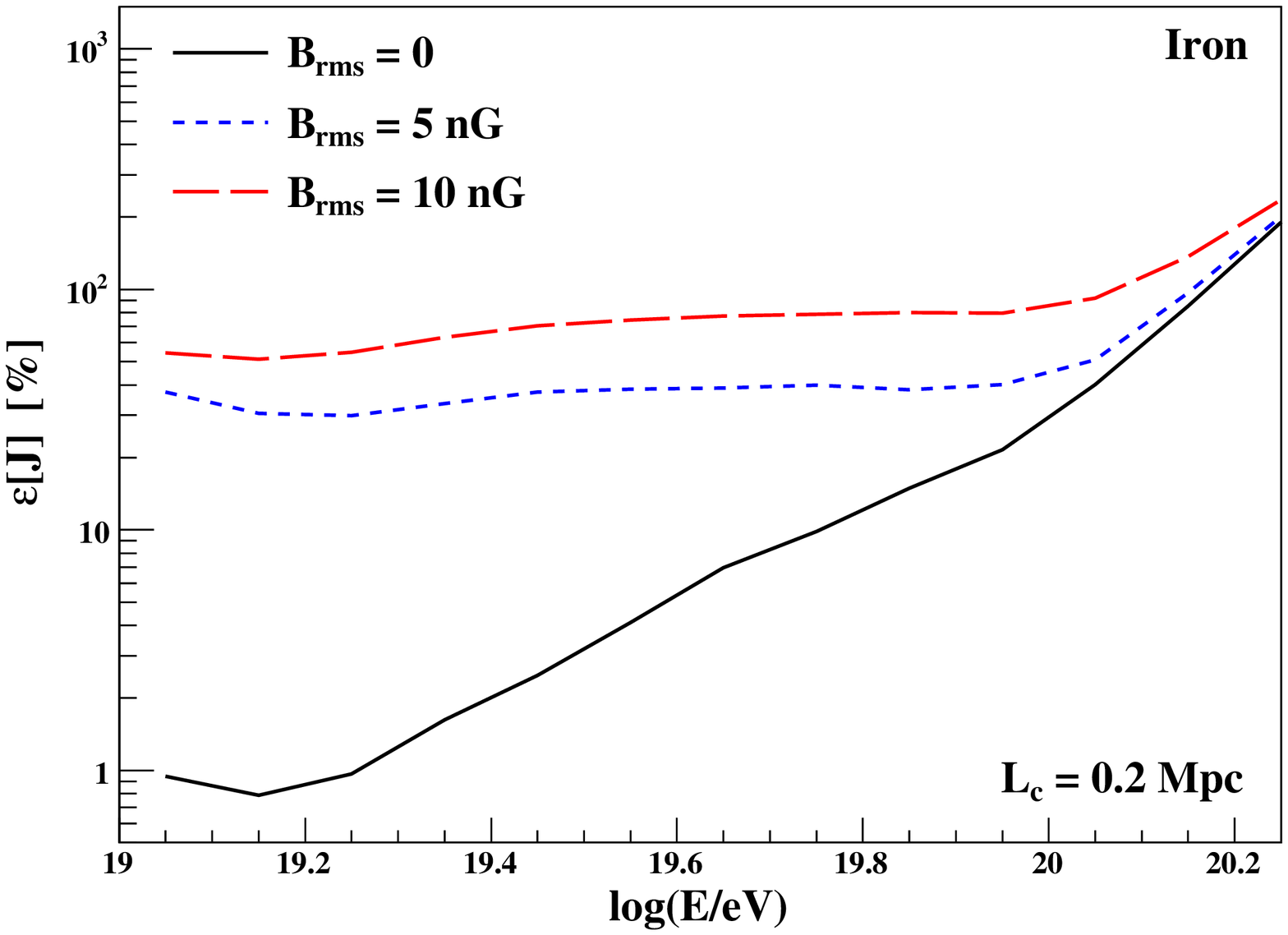}
\caption{\label{EpsRmin10} Relative standard deviation as a function of the logarithm of energy. The sources 
inject only protons (top panel) or iron nuclei (bottom panel). The IGMF is such that $B_{rms}=0$, $5$, and 
$10$ nG. The density of sources is $n_s=10^{-5}$ Mpc$^{-3}$ and $r_{min}=10$ Mpc. The spectral index is 
$\gamma=2.2$ for iron nuclei and $\gamma=2$ for protons, and $E_{cut}=10^{21}$ eV for both cases.}
\end{figure}

In Ref.~\cite{Ahlers:13}, it is shown that the relative standard deviation for a null IGMF is smaller for larger values 
of $r_{min}$. Figure \ref{EpsRmin10} shows the relative standard deviation for $r_{min}=10$ Mpc, density of sources 
$n=10^{-5}$ Mpc$^{-3}$, and for all configurations of the IGMF considered before. The top panel shows the results for 
protons and the bottom panel the ones for iron nuclei. The relative standard deviations are smaller compared with the 
$r_{min}=3$ Mpc case, for all configurations of the IGMF considered. The reason for that is that, as mentioned before, 
sources close to the observer introduce large ensemble fluctuations because the distribution of distances is $\propto r^2$. 
Therefore, when sources with $r \in [3, 10]$ Mpc are removed, the ensemble fluctuations decrease.    

\begin{figure}[!th]
\centering 
\includegraphics[width=.47\textwidth]{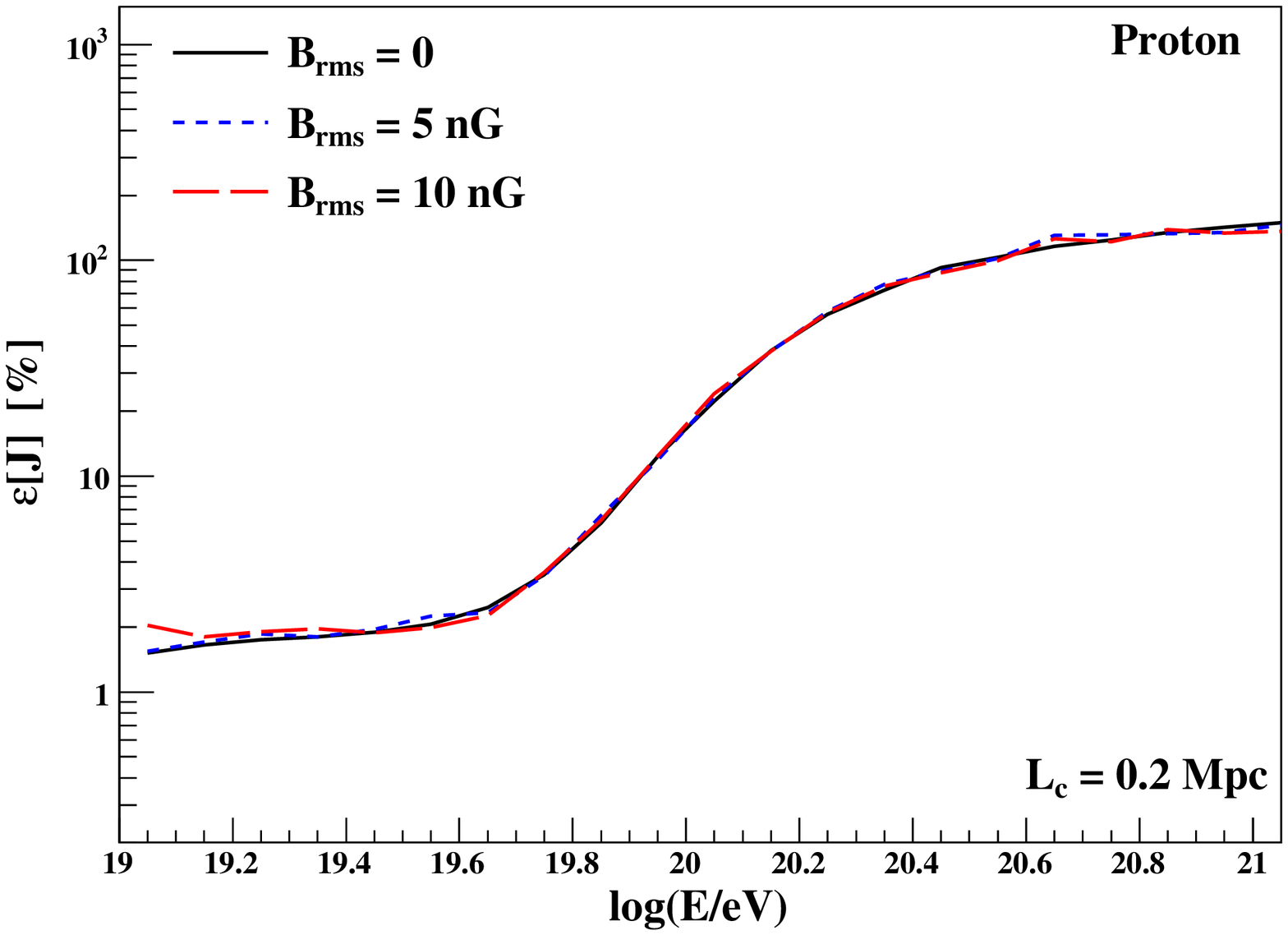}
\includegraphics[width=.47\textwidth]{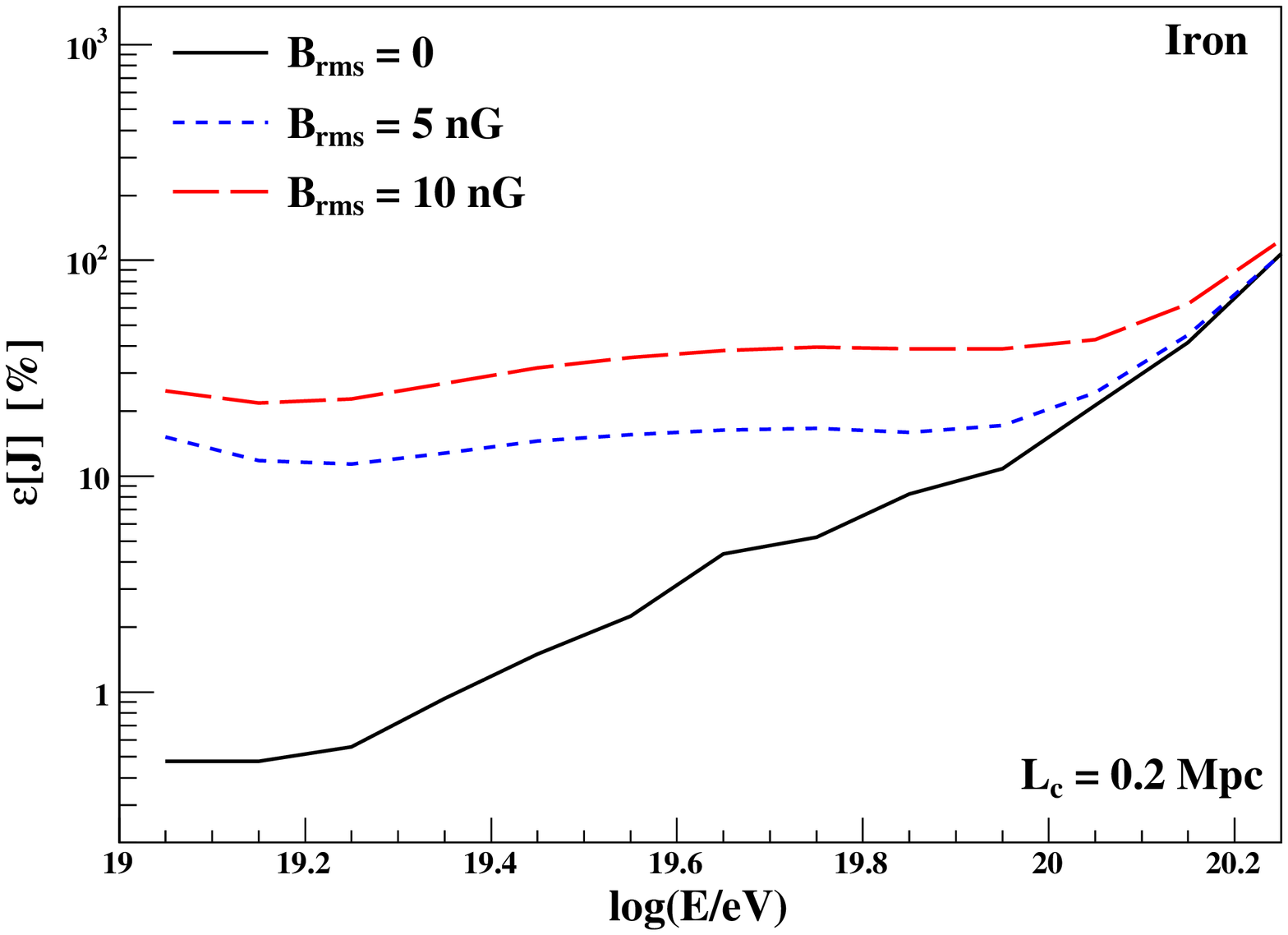}
\caption{\label{EpsN4} Relative standard deviation as a function of the logarithm of energy. The sources 
inject only protons (top panel) or iron nuclei (bottom panel). The IGMF is such that $B_{rms}=0$, $5$, and 
$10$ nG. The density of sources is $n_s=10^{-4}$ Mpc$^{-3}$ and $r_{min}=3$ Mpc. The spectral index is 
$\gamma=2.2$ for iron nuclei and $\gamma=2$ for protons, and $E_{cut}=10^{21}$ eV for both cases.}
\end{figure}

The ensemble fluctuations of both energy spectrum and composition profile depend on the density of sources. The larger 
the density of sources, the smaller the ensemble fluctuations. Figure \ref{EpsN4} shows the relative standard deviation 
for sources that inject only protons (top panel) or iron nuclei (bottom panel) for $n_s=10^{-4}$ Mpc$^{-3}$ and 
$r_{min}=3$ Mpc for both cases. By comparing with the bottom panels of Figs.~\ref{MEpsPr} and \ref{MEpsFe}, it is clear 
that the relative standard deviations decreases in all cases. The relative standard deviation for $n_s=10^{-5}$ Mpc$^{-3}$ 
is $\sim 30-35$ times larger than the ones corresponding to $n_s=10^{-4}$ Mpc$^{-3}$, for all cases considered. Note 
that for the cases corresponding to $r_{min}=10$ Mpc this ratio is in the reported range but slightly smaller than 
the one corresponding to the same cases but for $r_{min}=3$ Mpc.   

\section{Influence on the interpretation of the experimental data}
\label{Pheno}

\begin{figure}[!th]
\centering 
\includegraphics[width=.47\textwidth]{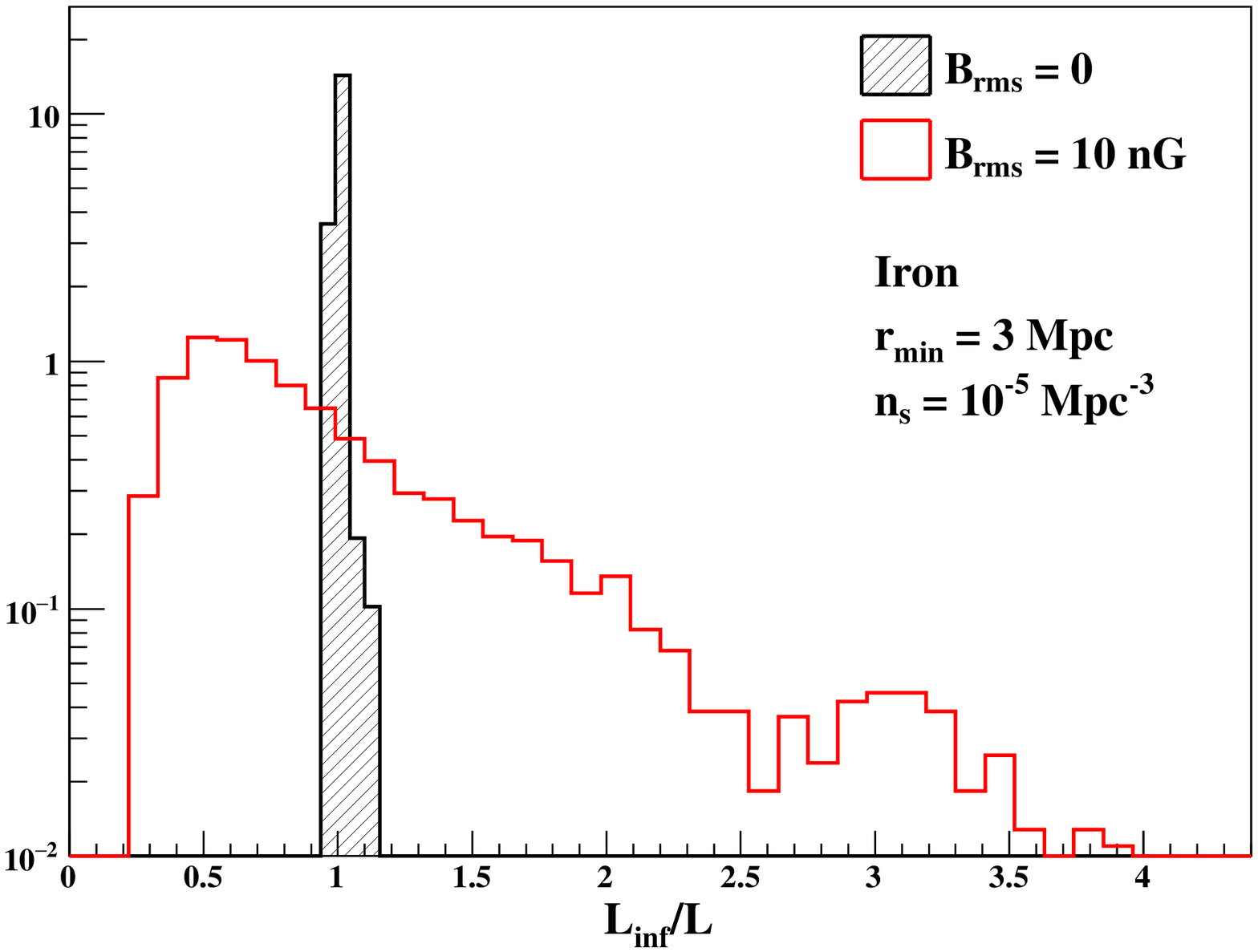}
\includegraphics[width=.47\textwidth]{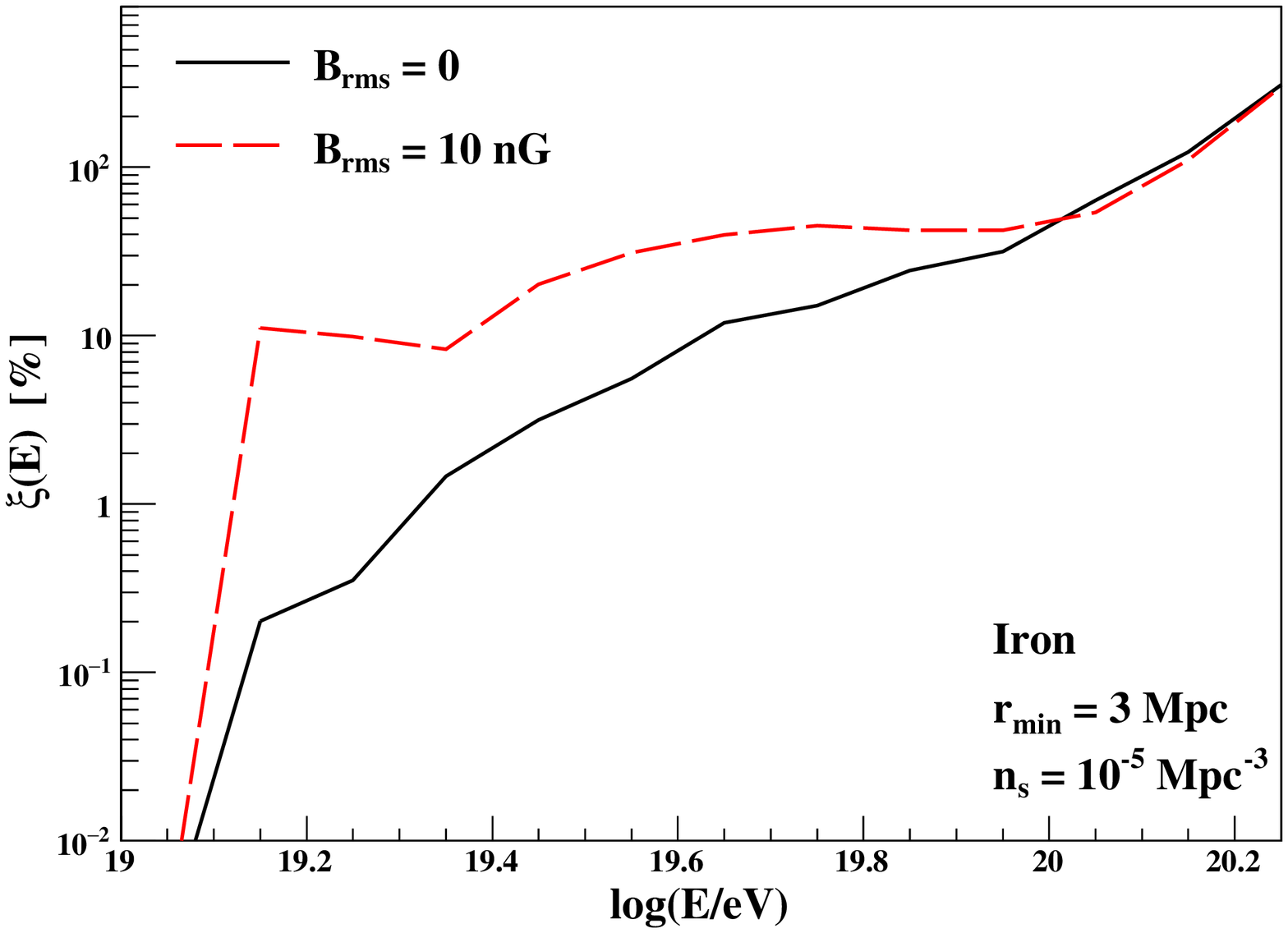}
\caption{\label{PhN5} Top panel: Distributions of the ratio between the inferred luminosity and the true one for sources 
injecting iron nuclei in the presence of a null and a non-null IGMF. The distributions are normalized to one. Bottom panel: 
$\xi$ as a function of the logarithm of energy. In this case, $n_s = 10^{-5}$ Mpc$^{-3}$ and $r_{min} = 3$ Mpc.}
\end{figure}

It is a common practice to fit the cosmic ray energy spectrum, obtained experimentally, with the mean value of the 
flux corresponding to a given astrophysical model. In order to study the effects of the IGMF on this type of practice, 
let us normalize the mean value of the flux, evaluated at the lowest value of energy considered ($E_0 = 10^{19.05}$ eV), 
to the one corresponding to a given realization of the source distribution evaluated at the same energy. The flux obtained
in this way is given by
\begin{equation}
\label{FitPh}
\tilde{J}(E)=\frac{J(E_0)}{E[J](E_0)}E[J](E),
\end{equation}
which is taken as an approximated representation of the real spectrum. Note that this procedure corresponds to change the 
luminosity of the sources; i.e., if $J(E_0)>E[J](E_0)$, a larger luminosity, compared with the true one, is inferred. The 
top panel of Fig.~\ref{PhN5} shows the distributions corresponding to the ratio between the inferred luminosity and the 
true one for sources injecting iron nuclei with the spectral index and cutoff energy considered before, $r_{min} = 3$ Mpc, 
and $n_s = 10^{-5}$ Mpc$^{-3}$. The IGMF is such that $B_{rms}=0$ and $B_{rms}=10$ nG. From the figure, it is easy to see 
that for the non-null IGMF case the inferred luminosity can be $\sim 4$ times larger or smaller than the true one. In contrast, 
for the case of a null IGMF the inferred luminosity differs from the true one in less than $\sim 15$\%. The increase of the 
error on the inferred luminosity for a non null IGMF is due to the larger ensemble fluctuations compared with the ones 
corresponding to a null IGMF.           

In order to estimate the relative error corresponding to the use of the mean value of the flux to represent the energy 
spectrum, obtained for a particular realization of the position of the sources, the following quantity is introduced
\begin{equation}
\label{XiDef}
\xi(E)=\frac{\sigma[\Delta J](E)}{E[J](E)},
\end{equation}
where $\Delta J(E)=J(E)-\tilde{J}(E)$. Note that, by definition, $\xi(E_0)=0$. The bottom panel of Fig.~\ref{PhN5} shows $\xi$ 
as a function of the logarithm of the energy. It can be seen that, for energies smaller than $10^{20}$ eV, the relative error 
corresponding to a non-null IGMF case is larger than the one corresponding to the null IGMF case. The difference is larger for 
energies of the order of $10^{19.2}$ eV. For energies larger than $10^{20}$ eV, the differences are small.  

\begin{figure}[!th]
\centering 
\includegraphics[width=.47\textwidth]{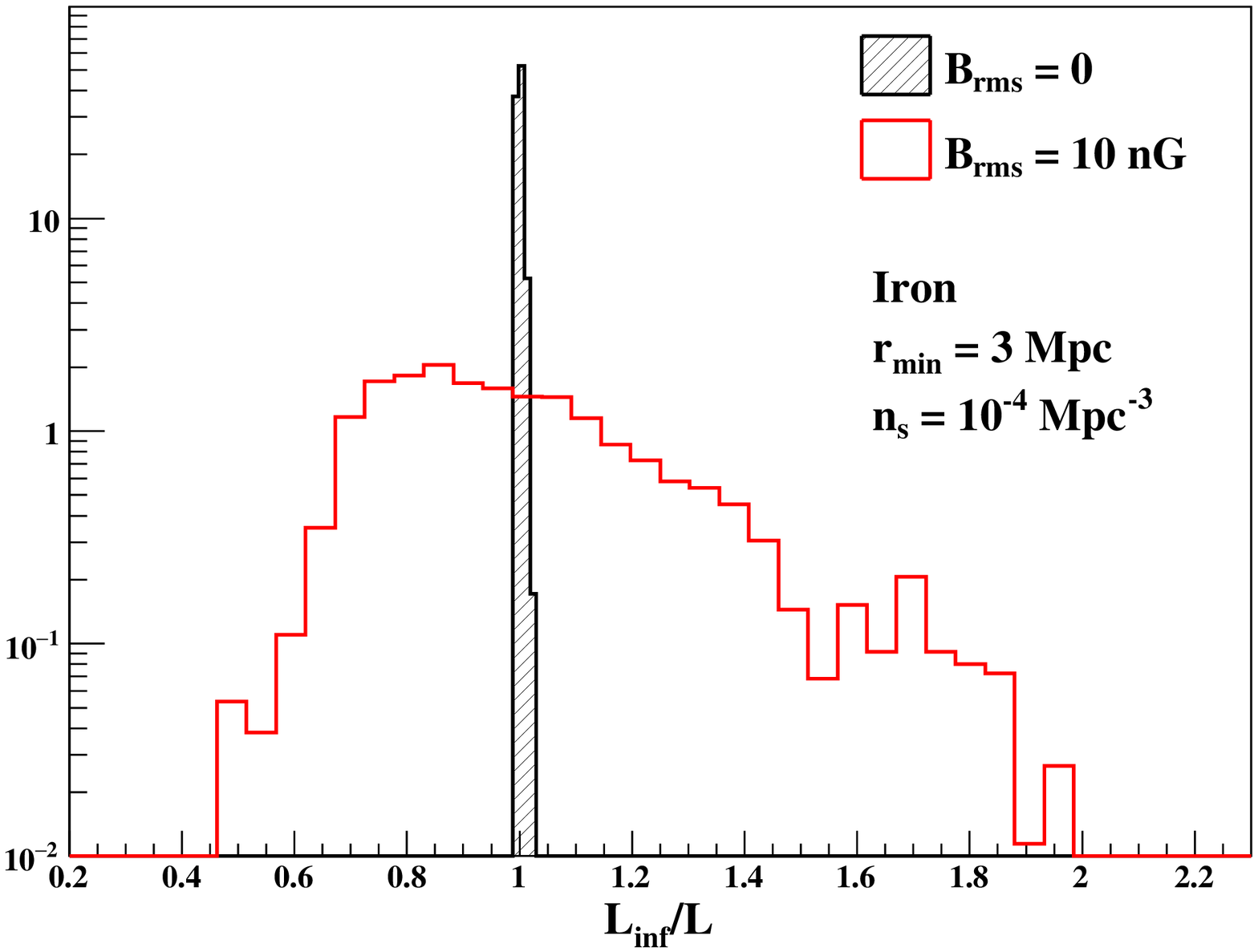}
\includegraphics[width=.47\textwidth]{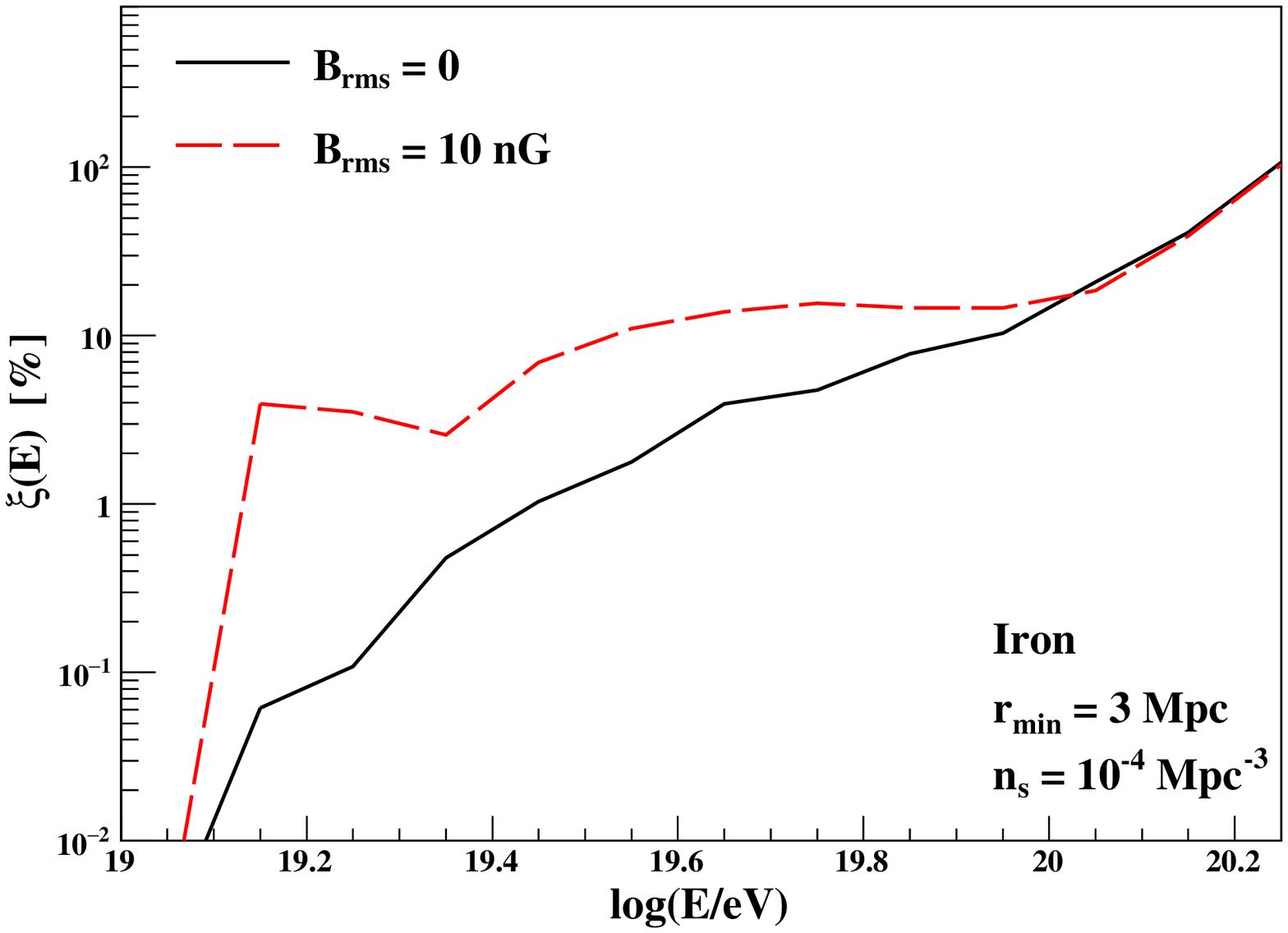}
\caption{\label{PhN4} Top panel: Distributions of the ratio between the inferred luminosity and the true one for sources 
injecting iron nuclei in the presence of a null and a non-null IGMF. The distributions are normalized to one. Bottom panel: 
$\xi$ as a function of the logarithm of energy. In this case, $n_s = 10^{-4}$ Mpc$^{-3}$ and $r_{min} = 3$ Mpc.}
\end{figure}

Figure \ref{PhN4} shows the results obtained by increasing the density of sources to $n_s = 10^{-4}$ Mpc$^{-3}$. From the top panel 
of the figure, it can be seen that, in this case, the inferred luminosity can be $\sim 2$ times larger or smaller than the true one. 
For the case of a null IGMF the inferred luminosity  differs from the true one by less than $\sim 3$\%. As expected, the error on 
the determination of the luminosity is reduced, but it is still quite large for the case of a non-null IGMF. 
From the bottom panel of Fig.~\ref{PhN4}, it can be seen that $\xi$ is smaller than in the previous case for both scenarios
considered. However, also in this case the difference between $\xi$ for a null and a non-null IGMF is still quite large 
in the region of $10^{19.2}$ eV. As in the previous case, the difference is small for energies larger than $10^{20}$ eV.

\section{Prospect of observation}
\label{POJE}

The study of the ensemble fluctuations requires a large statistics at the highest cosmic ray energies. The Extreme 
Universe Space Observatory on board the Japanese Experimental Module (JEM-EUSO) is ideal for this type of study 
as shown in Ref.~\cite{AlhersICRC:13}. JEM-EUSO will consist of the installation of a fluorescence telescope on the 
International Space Station pointing towards Earth's surface \cite{JEMEUSO}. The atmospheric air showers 
originated by the interaction of the cosmic rays with the air molecules generate fluorescence and Cherenkov light. 
A fraction of this light can be collected by the telescope, allowing for the reconstruction of the main shower 
parameters (arrival direction, primary energy, etc.). The telescope can be operated in nadir mode (pointing in 
the nadir direction) or tilted mode (forming an angle with the nadir direction). The exposure in tilted mode 
increases quite fast with the nadir angle, but the energy threshold also increases. Note that the annual exposure 
of JEM-EUSO in one year of data taking in Nadir mode is $9-10$ times larger than that of The Pierre Auger 
Observatory \cite{Adams:13}. This will allow the collection of a cosmic ray sample of unprecedented statistics at 
the highest energies.

\begin{figure*}[!th]
\centering 
\includegraphics[width=.47\textwidth]{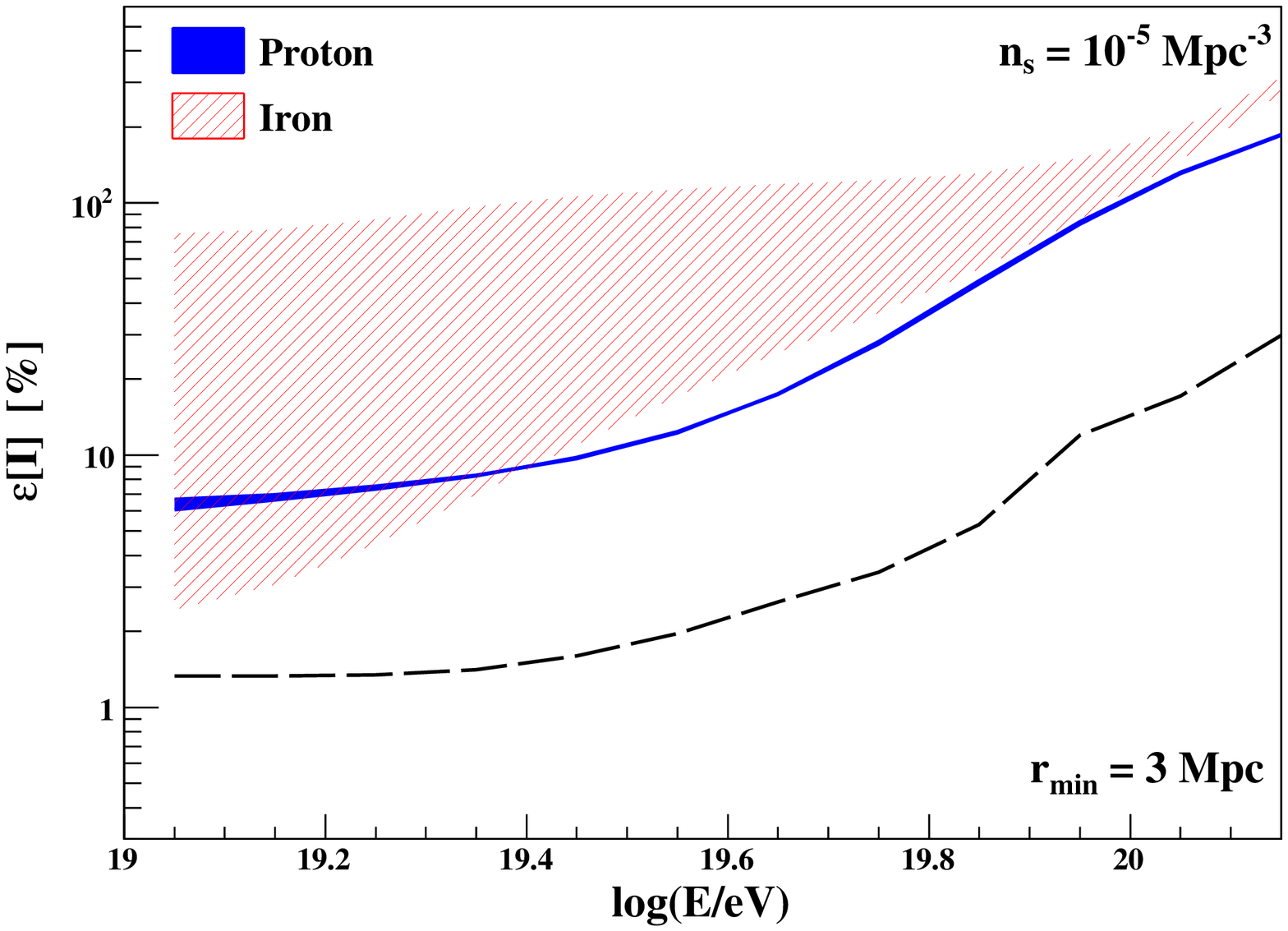}
\includegraphics[width=.47\textwidth]{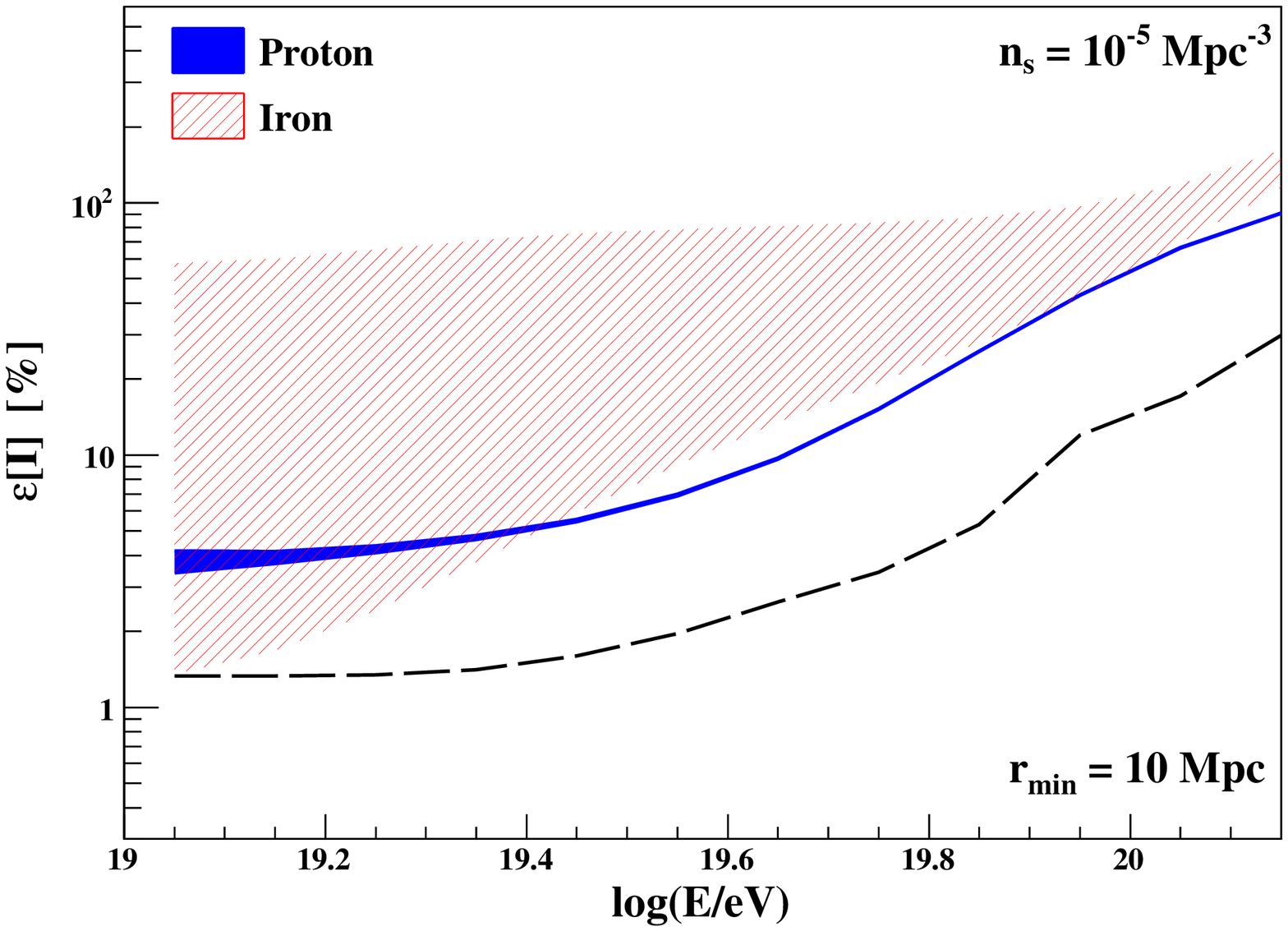}
\includegraphics[width=.47\textwidth]{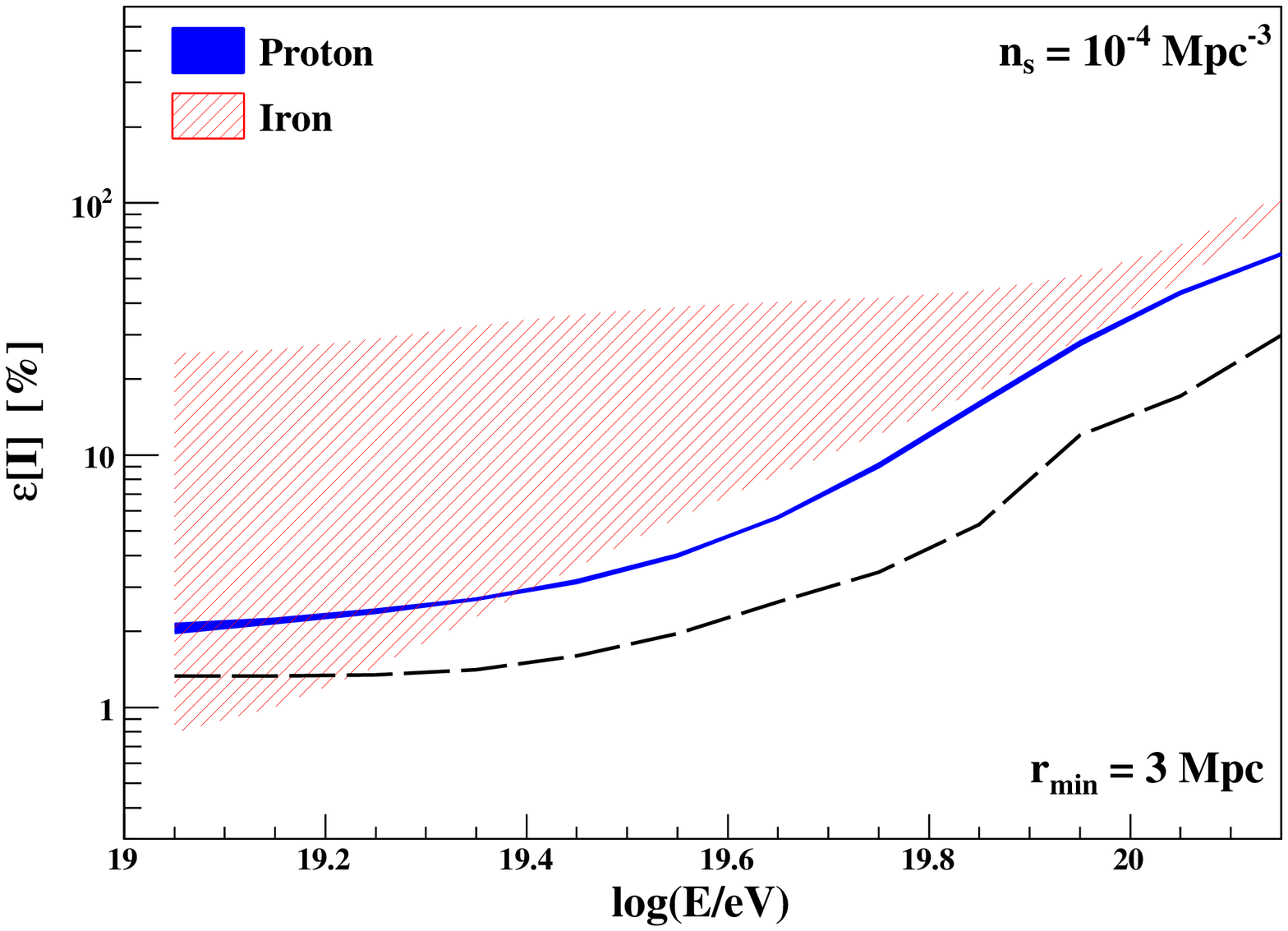}
\includegraphics[width=.47\textwidth]{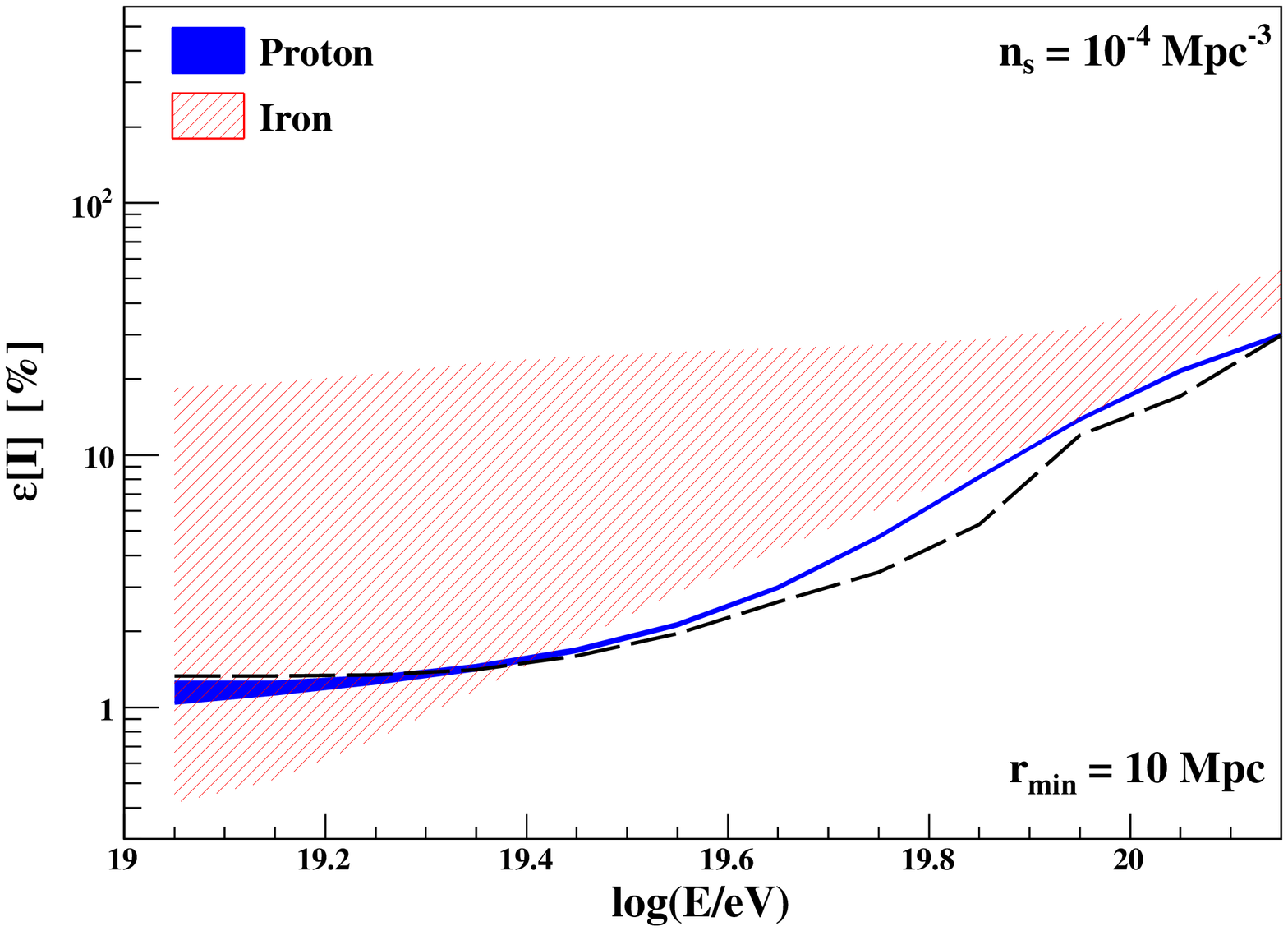}
\caption{\label{JE} Relative standard deviation of the integral cosmic ray energy spectrum as a function 
of the logarithm of energy. Solid dashed line: Statistical fluctuations of the energy spectrum for 5 years of 
data taking of JEM-EUSO in nadir mode.}
\end{figure*}

It is worth mentioning that the integral cosmic ray energy spectrum, defined as 
\begin{equation}
I(E)=\int_E^\infty dE' J(E'),
\end{equation}
also presents ensemble fluctuations. Figure \ref{JE} shows the relative standard deviation of the integral energy
spectrum compared with the statistical fluctuations of the integral energy spectrum for 5 years of data taking of 
JEM-EUSO in nadir mode. For the latter, Poissonian fluctuations are assumed. Also the energy spectrum obtained 
by The Pierre Auger Observatory \cite{Auger:11} and the JEM-EUSO exposure calculated in Ref.~\cite{Adams:13} are used. 
The calculation is done for the two models considered before in which the sources inject only protons or iron nuclei. 
The shadowed regions represent the range of the relative standard deviation of the integral spectrum for 
$B_{rms} \in [0,10]$ nG. The top panels of the figure correspond to $n_s=10^{-5}$ Mpc$^{-3}$, and the bottom panels 
correspond to $n_s=10^{-4}$ Mpc$^{-3}$. Also, the left panels correspond to $r_{min} = 3$ Mpc, and the right panels 
correspond to $r_{min} = 10$ Mpc.    

From the figure, it can be seen that for $n_s=10^{-5}$ Mpc$^{-3}$ the ensemble fluctuations are larger than the Poissonian 
fluctuations, in the whole energy range, for both models considered. For $n_s=10^{-4}$ Mpc$^{-3}$, $r_{min} = 3$, and 
protons, also the ensemble fluctuations are larger than the Poissonian fluctuations, in the whole energy range considered. 
However, for iron nuclei and small values of the IGMF intensity, the ensemble fluctuations are larger than the Poissonian 
fluctuations only if $E>10^{19.3}$ eV. For $n_s=10^{-4}$ Mpc$^{-3}$, $r_{min} = 10$, and protons, the Poissonian fluctuations 
are of the order of or even larger than the ensemble fluctuations in the whole energy range. However, for iron nuclei and 
small values of the IGMF intensity, the ensemble fluctuation are larger than the Poissonian fluctuations for $E>10^{19.5}$ eV.
Note that for the iron injection model and $B_{rms} \gtrsim 5$ nG the ensemble fluctuations are larger than the Poissonian 
fluctuations for all cases considered. Therefore, no null IGMF and heavy primaries at the highest energies favor the 
observation of the ensemble fluctuations with JEM-EUSO. Also, it is possible that the lifetime of the JEM-EUSO mission may be 
extended to more than 5 years. It will be important for the observation of the ensemble fluctuations.            

\section{Conclusions}
\label{conc}

In this work, the influence of a turbulent IGMF on the ensemble fluctuations of the cosmic ray energy
spectrum and composition profile has been studied. Sources injecting only protons and only iron nuclei 
have been considered. Turbulent IGMFs of coherence length $L_c \cong 0.2$ Mpc and intensities $B_{rms}=5$ 
and 10 nG have been taken into account. It has been found that the influence of such magnetic fields on 
the ensemble fluctuations in the proton model is almost negligible. For the case of iron nuclei, it has 
been found that, at energies of the order of a few $10^{19}$ eV and for $B_{rms} \geq 5$ nG, the ensemble 
fluctuations are more than one order of magnitude larger compared with the case of a null IGMF. This 
difference decreases with increasing energies due to the fact that cosmic rays of higher energies are 
less affected by the same magnetic field configuration. The same behavior has been found for the ensemble 
fluctuations of the composition profile. Also, the composition becomes lighter for increasing values of 
the IGMF intensity. As in the case of the ensemble fluctuations of the energy spectrum, the differences 
with the null IGMF case decrease for increasing values of energy. The increase of the ensemble fluctuations 
can be understood from the fact that cosmic rays propagating in a non-null IGMF are injected by sources that 
are closer. This is caused by the increase of the path length of the particles combined with their 
interactions with the low energy photons of the radiation field present in the intergalactic medium. 
The lighter composition can be explained by the increase of the path length traveled by the particles 
in the presence of a non-null IGMF.

It has been found that, for sources injecting heavy primaries, the presence of a non-null IGMF largely 
increases the uncertainty on the inferred luminosity of the sources, when the mean value of the flux 
is used to represent the energy spectrum corresponding to a particular realization of the position
of the sources. Also, the uncertainty on the flux shape is in some energy regions much larger for 
the case of a non-null IGMF. These uncertainties become smaller for increasing values of the density
of sources. However, for the cosmic ray and IGMF models considered in this work, even for densities 
of the sources of the order of $n_s=10^{-4}$ Mpc$^{-3}$, these uncertainties are quite large.   

Regarding the possible observation with the next generation of cosmic ray observatories like JEM-EUSO, 
it has been demonstrated that the ensemble fluctuations on the integral energy spectrum will be easier 
to be observed for models including the injection of heavy nuclei combined with a non-null IGMF. In any
case, at least for the models considered in this work, it seems possible to observe ensemble fluctuations,
even for density of sources as large as $10^{-4}$ Mpc$^{-3}$, in 5 years of observation with the JEM-EUSO
telescope in nadir mode. Note that the extension of the lifetime of the JEM-EUSO mission will be of great
benefit for the observation of the ensemble fluctuations.             

\begin{acknowledgments}
A.D.S.~is a  member of the Carrera del Investigador Cient\'ifico of CONICET, Argentina. This work is supported
by CONICET PIP 114-201101-00360 and ANPCyT PICT-2011-2223, Argentina. This work was partially supported by
PAPIIT-UNAM, Red FAE, and Red CyTE from CONACyT M\'exico.
\end{acknowledgments}

\appendix

\section{Comparison with other calculations}
\label{App}

\begin{figure}[!th]
\centering 
\includegraphics[width=.47\textwidth]{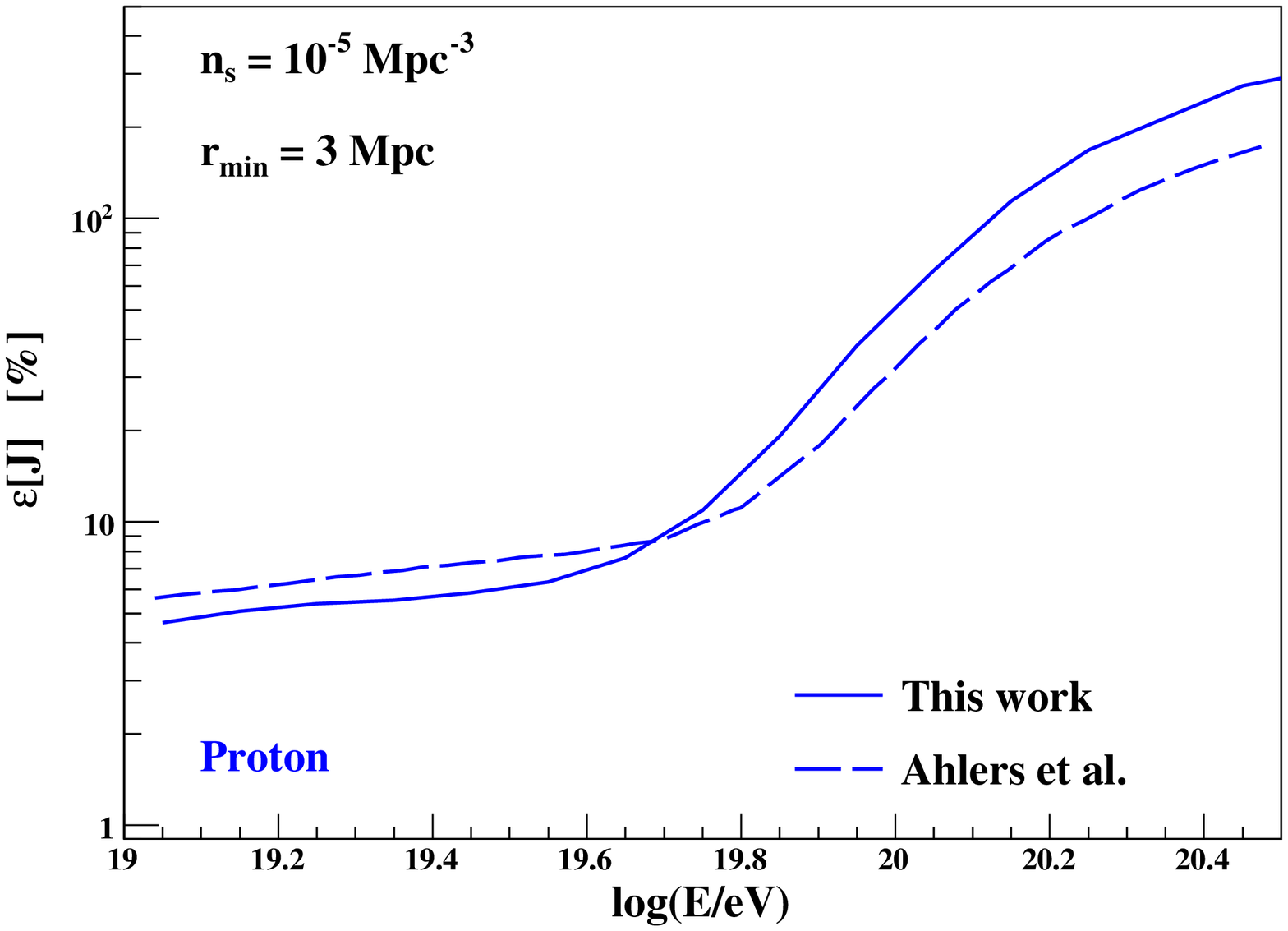}
\includegraphics[width=.47\textwidth]{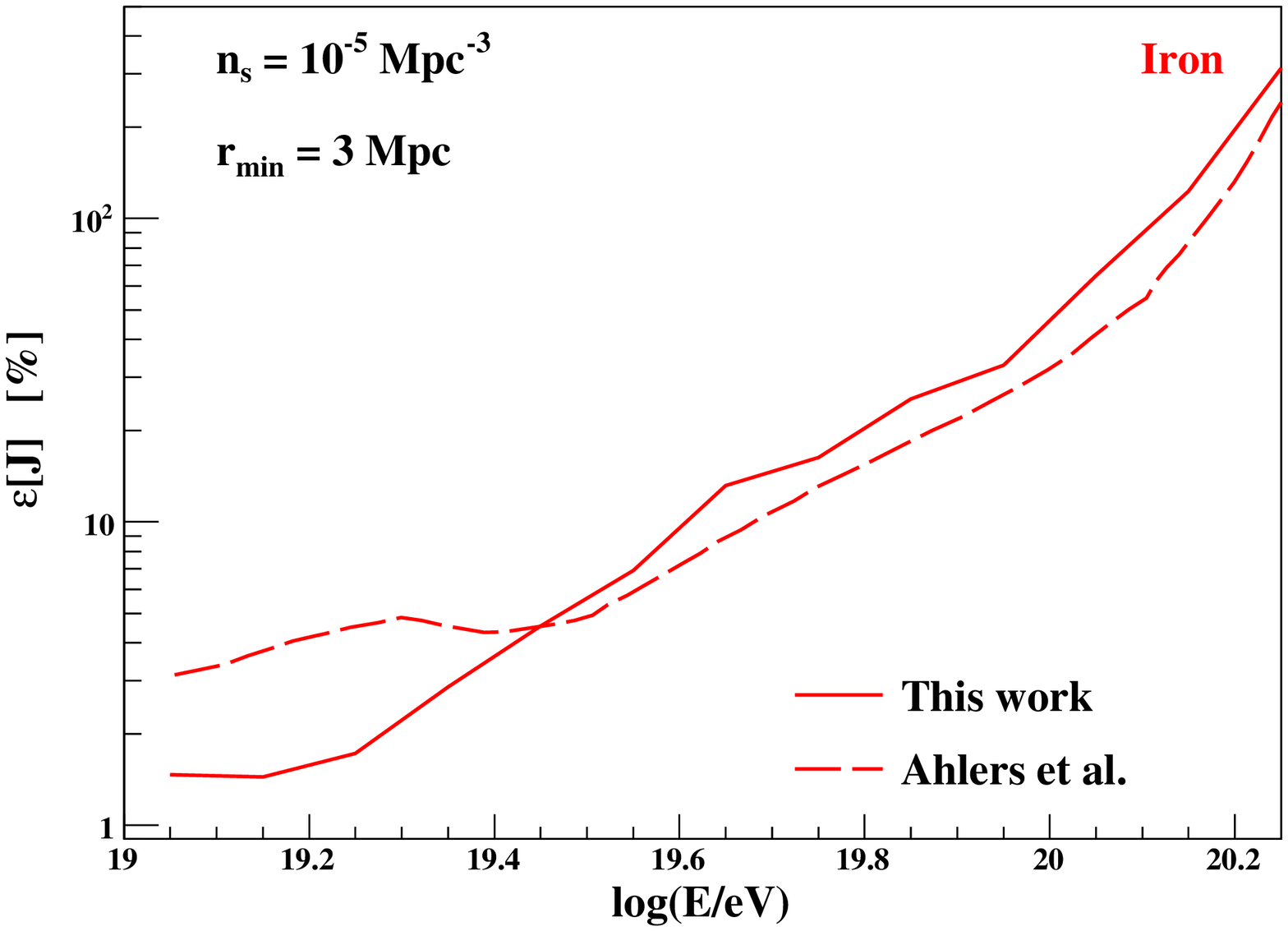}
\caption{\label{Comp} Relative standard deviation as a function of the logarithm of energy obtained in this work (solid 
lines) and in Refs.~\cite{Ahlers:13,AlhersICRC:13} (dashed lines), corresponding to  protons (top panel) and iron nuclei 
(bottom panel).}
\end{figure}

Figure \ref{Comp} shows the comparisons corresponding to protons (top panel) and iron nuclei (bottom panel) between 
the relative standard deviation $\varepsilon[J]$ obtained in this work and in Refs.~\cite{Ahlers:13,AlhersICRC:13}.
The calculations correspond to a null IGMF, density of sources $n_s = 10^{-5}$ Mpc$^{-3}$, and $r_{min}=3$ Mpc. 
The spectral indexes are $\gamma=2.2$ and $\gamma=2$ for protons and iron nuclei, respectively. The cutoff energy is
$E_{cut} = 10^{21}$ eV in both cases.  

From the figure, it can be seen that, although the curves obtained in this work and the ones from 
Ref.~\cite{Ahlers:13,AlhersICRC:13} have a similar trend, there are differences. In particular, for protons at low 
energies, $E \in [10^{19}, 10^{19.8}]$ eV, both calculations differ in less than $\sim 18 \%$ and for energies larger 
than $10^{19.8}$ eV the difference between both calculations increases reaching values of $\sim 70 \%$. For the case of 
iron nuclei, the differences reach values of at most $\sim 60 \%$ in the energy range considered. As can be seen from 
the bottom panel of the figure, the larger differences are given at lower energies ($E\lesssim 10^{19.4}$ eV).

\end{document}